\documentclass[pre,onecolumn,showpacs,superscriptaddress,aps]{revtex4}

\usepackage[toc,page,title,titletoc,header]{appendix}
\usepackage[usenames]{color}
\usepackage{amsfonts,amssymb,latexsym,amsmath}
\usepackage{amsthm}
\usepackage{graphicx}
\usepackage{float}
\usepackage{dcolumn}
\usepackage{times}
\usepackage{ulem}
\usepackage[hang,nooneline]{subfigure}
\usepackage{epstopdf}

\newcounter{fig}   

\newcommand{\beq}{\begin{equation}}
\newcommand{\eeq}{\end{equation}}
\newcommand{\eps}{\epsilon}

\begin{document}

\title{Lattices With Internal Resonator Defects}
\author{S.~Hauver}
\affiliation{Department of Mathematics and Statistics, University of Massachusetts, Amherst, MA 01003-4515, USA}
\author{X.~He}
\affiliation{Department of Electrical and Computer Engineering, University of Massachusetts, Amherst, MA 01003-4515, USA}
\author{D. Mei}
\affiliation{Department of Mathematics and Statistics, University of Massachusetts, Amherst, MA 01003-4515, USA}
\author{E.G. Charalampidis}
\affiliation{Department of Mathematics and Statistics, University of Massachusetts, Amherst, MA 01003-4515, USA}
\author{P.G. Kevrekidis}
\affiliation{Department of Mathematics and Statistics, University of Massachusetts, Amherst, MA 01003-4515, USA}
\author{E.~Kim}
\affiliation{Department of Aeronautics and Astronautics, University of Washington, Seattle, WA 98195-2400}
\affiliation{Automotive Hi-Technology Research Center, Division of Mechanical System Engineering, Chonbuk National University, Jeonju-si, Jeollabuk-do, 54896, Republic of Korea}
\author{J.~Yang}
\affiliation{Department of Aeronautics and Astronautics, University of Washington, Seattle, WA 98195-2400}
\author{A. Vainchtein}
\affiliation{Department of Mathematics, University of Pittsburgh, Pittsburgh, PA 15260, USA}

\begin{abstract}

We consider a variety of settings involving chains with one or more defects stemming
from the introduction of nodes bearing internal resonators. Motivated by experimental
results in woodpile elastic lattices with one or two defects, we
consider a variety of different theoretical scenarios. These include multi-defect chains
and their ability to transmit, reflect, and especially trap energy; they also include
settings with linear vs. nonlinear defects of variable interaction exponent. Moreover,
they involve defects which are spatially separated and either statically, or more effectively
dynamically, enable the confinement of energy between the separated defects. Wherever
possible, comparisons of the experiments with numerical simulations,
as well as with theoretical intuition
are also offered, to provide a justification for the
observed findings.
\end{abstract}

\date{\today}
\maketitle

\section{Introduction}

The study of granular crystals and of related topics has received considerable attention
especially over the last decade. To a considerable extent this is
arguably due to the significant experimental progress that has complemented theoretical
and numerical investigations; see, e.g.,~\cite{Nesterenko2001,Sen2008,pgk_review,Theocharis_rev} for earlier
and~\cite{physicstod,vakbook,chong_rev} for recent
reviews. In
these media with extensively tunable properties, traveling waves have been of particular
interest. More recently numerous other excitations have been examined including, but not
limited to, defect modes, bright and dark breathers, and shock waves~\cite{physicstod,chong_rev}.
At the same time, a diverse host of applications including, e.g., actuating devices~\cite{dev08},
acoustic lenses~\cite{Spadoni}, mechanical diodes~\cite{china1,china2,Nature11}, logic
gates~\cite{Feng_transistor2014} and sound scramblers \cite{dar05,Nesterenko2005} has
also been proposed, adding a more practical dimension to the theoretical appeal of the
subject.

A twist on this theme of granular crystals that has led to numerous recent studies concerns
the subject of the so-called locally resonant granular crystals, otherwise known as mass-in-mass
(MiM) or mass-with-mass (MwM) systems. The MiM and MwM systems have already been experimentally
realized in~\cite{bonanomi} and~\cite{gatz}, respectively. These realizations were chiefly
linear, due to externally imposed static precompression of the chain, and geared towards the
remarkable metamaterial-type properties that these systems possess. A prototype of a strongly
nonlinear granular chain with a single MwM defect was also demonstrated in~\cite{annavain},
where numerical investigations of the system, complemented by multiscale asymptotic analysis of
a reduced model, demonstrated an ability of such defect to trap and reflect portions of the
energy carried by a solitary wave. More recently, a different type of experiment was realized
showcasing highly nonlinear propagation in a locally resonant granular system~\cite{eunho02}.
In particular, this experiment featured a so-called woodpile configuration consisting of
orthogonally-stacked rods~\cite{eunho01} and demonstrated that the bending vibrations of the
rods can play the role of the local resonator within the chain. It was also shown that depending
on the properties of the system (i.e., the length of the rods), one can controllably incorporate
one or more such resonators and observe unique types of waveforms not previously explored
in granular chains, including weakly nonlocal solitary waves.

In the present work, we consider a strongly nonlinear granular chain with a finite number of MwM
defects, focusing particularly on the cases of adjacent and separated defects that were only briefly
explored in \cite{annavain}. This setting interpolates between the single-defect case that was the
main focus of~\cite{annavain} and the case of a woodpile lattice of~\cite{eunho01,eunho02} where
each granule is effectively coupled to a local resonator. To motivate this work, we begin by presenting
experimental results for woodpile lattices involving one and two defects represented by longer rods.
These experiments allow us to infer the fraction of transmitted, reflected, as well as trapped kinetic
energy for each case, in very good agreement with the corresponding simulation results. In light of these
experimental possibilities, we theoretically explore a number of variants of the problem. More specifically,
we consider a ``defective'' region of variable domain and examine
how the different energy fractions scale with the size of this region. We also examine the role of a
nonlinear vs. a linear resonator coupling, by varying the exponent in the resonator term. We explore
the possibility of separating the two defects, and also of using such a separation to attempt to induce
(possibly also dynamically) a trapping of the
traveling solitary wave between the two defects. We offer detailed comparisons of our experimental
results with corresponding numerical simulations, as well as, wherever possible, of the numerical
computations with theoretical considerations.

Our presentation is structured as follows. In Section~II, we introduce the model. In Section~III, we
give an overview of our corresponding experimental results and their comparison to numerical computations.
In Sections~IV (for single or adjacent defects) and V (for more general,
and also dynamically variable in time, settings)
we explore a number of scenarios (as discussed above) via direct numerical simulations
and corresponding theoretical analysis. Finally, in Section~VI, we summarize our findings and present
our conclusions.

\section{The model}

\begin{figure}
\centering
\includegraphics[scale=0.5]{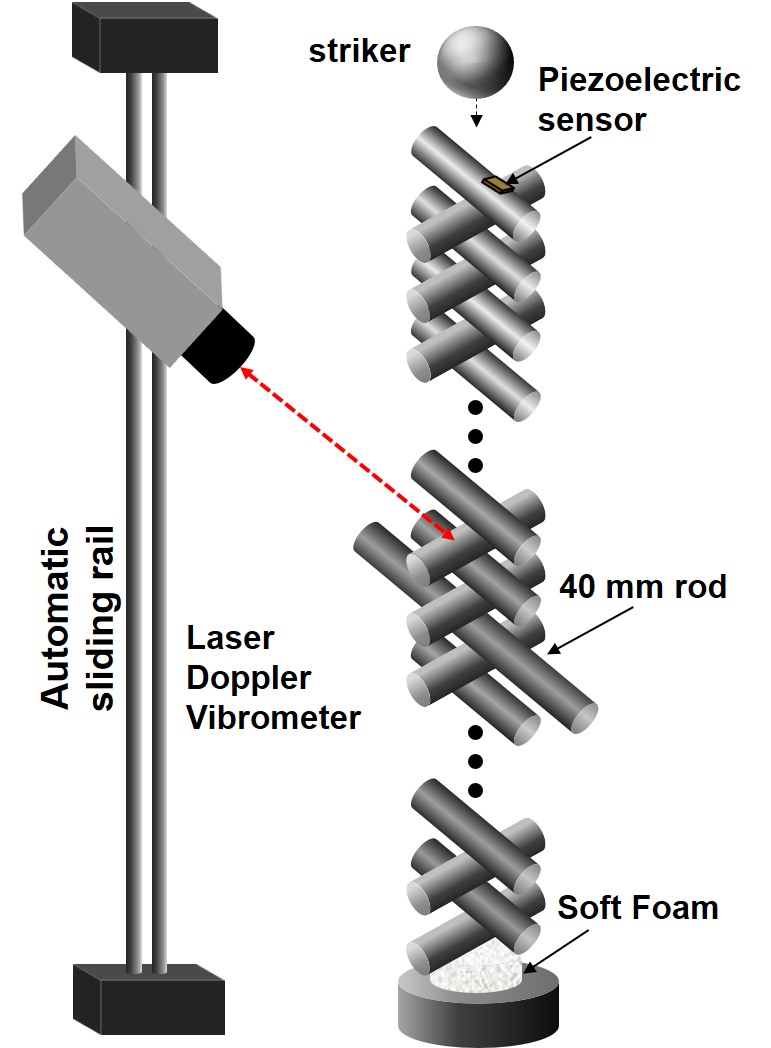}
\caption{Experimental setup involving the woodpile chain, the striker leading
to the formation of the wave and the laser Doppler vibrometer enabling
its probing.}
\label{fig:expSetup}
\end{figure}

We begin by considering a nonlinear granular chain with a single linear internal resonator
defect. Mathematically such a system is represented by
\beq
\begin{split}
M\ddot{u}_n&=a\biggl(\left[u_{n-1}-u_n\right]_{+}^{3/2}%
-\left[u_{n}-u_{n+1}\right]_{+}^{3/2}\biggr)-K(u_0-v_0)\delta_{n0}, %
\quad n=-N,\dots,N,\\
m\ddot{v}_0&=K(u_0-v_0),
\end{split}
\label{eq:orig_eqns}
\eeq
where $\delta_{n0}$ is the Kronecker delta, i.e., $\delta_{n0}=1$ when $n=0$ and zero otherwise.
Under certain conditions, these equations describe the effective dynamics of a woodpile chain
of orthogonally stacked cylindrical rods shown in Figure~\ref{fig:expSetup}, where all rods are
of the same length and mass $M$ except for a single ``defect" rod in the middle, which is longer.
In Eq.~\eqref{eq:orig_eqns} this rod corresponds to $n=0$, and the displacement of $n$th rod is
denoted by $u_n(t)$, with $\ddot{u}_n(t) = d^2 u_n(t) / dt^2$. The rods interact via Hertzian
contact forces characterized by the exponent of $3/2$ and constant $a>0$ that depends on the
material and geometry of the rod; {its explicit form is given by $a=(2E\sqrt{R})/(3(1-\nu^{2}))$,
where $E$ is the Young's modulus, $R$ the radius of each cylindrical rod and $\nu$ is the Poisson's
ratio.} The tensionless character is indicated by the subscript $+$, which means that the corresponding term is nonzero
only when the quantity in the parentheses is positive. As explained
in more detail below (see, also~\cite{eunho01,eunho02}), the first bending mode of the defect rod
at $n=0$ can be described by a linear internal resonator, where an effective mass $m$ with displacement
$v_0$ is attached to the primary mass $M$ by a linear spring of stiffness $K>0$. Here we assume a
parameter range in which other vibration modes of the defect rod as well as internal vibrations
of the shorter rods can be neglected along with possible sources of energy dissipation.

An impact applied to the chain, i.e., a striker, generates a solitary wave that eventually interacts
with the defect. To model this, we will consider the initial conditions
\beq
u_{n}(0)=v_{0}(0)=0, \quad \dot{u}_{n}(0)=\dot{v}_{0}(0)=0, \quad n \neq n_\text{S}, \quad \dot{u}_{n_\text{S}}(0) = V,
\label{eq:orig_ics}
\eeq
where the initially excited $n_\text{S}$th site above the defect ($n_\text{S}<0$) has impact velocity $V>0$.
Here and in what follows, $n$ ranges from $-N$ to $N$.

For the analysis purposes it is convenient to rescale the governing equations~\eqref{eq:orig_eqns}
and the initial conditions~\eqref{eq:orig_ics} by introducing dimensionless displacements
$\bar{u}_n$, $\bar{v}_0$ and time $\bar{t}$ related to the original variables via~\cite{annavain,linden}
\[
u_n=\biggl(\dfrac{M V^2}{a}\biggr)^{2/5}\bar{u}_n, \quad v_0=\biggl(\dfrac{M V^2}{a}\biggr)^{2/5}\bar{v}_0,
\quad
t=\dfrac{1}{V}\biggl(\dfrac{MV^2}{a}\biggr)^{2/5}\bar{t}.
\]

The two dimensionless parameters are the ratio of the two masses
\beq
\eps=\dfrac{m}{M},
\label{eq:eps}
\eeq
and the strength of the linear coupling measured by $\kappa=K/(M^{1/5}a^{4/5}V^{2/5})$.
Dropping the bars on the new variables, we obtain
\beq
\begin{split}
\ddot{u}_n &=\left[u_{n-1}-u_n\right]^{3/2}_+ - \left[u_n - u_{n+1}\right]^{3/2}_+ %
- \kappa(u_0-v_0) \delta_{n0},
\\
\eps\ddot{v}_0 &= \kappa(u_0 - v_0),
\end{split}
\label{eq:rescaled_eqns}
\eeq
and
\beq
u_n(0)=v_0(0)=0, \qquad \dot{v}_0(0)=0=\dot{u}_n(0), \quad n \neq n_\text{S}, \qquad \dot{u}_{n_\text{S}}(0)=1.
\label{eq:rescaled_ICs}
\eeq

An important diagnostic quantity of the conservative system under consideration is the total
energy
\begin{equation}
E = \sum_{n=-N}^N e_n,
\end{equation}
where $e_{n}$ stands for the energy density given by
\begin{equation}
e_n =  \frac{1}{2}\dot{u}_n^2 + \frac{\epsilon}{2}\dot{v}_0^2\delta_{n0} +%
\frac{\kappa}{2}(u_0 - v_0)^2\delta_{n0}
+\frac{1}{5}\Big\lbrace\left[u_{n-1}-u_n\right]_+^{5/2}+\left[u_n - u_{n+1}\right]_+^{5/2}\Big\rbrace.
\end{equation}
This will also enable us to characterize the fractions of the energy that
will be reflected and transmitted from, as well as trapped inside  the
``defective'' region. The corresponding fractions of the energy can be
defined as follows:
\beq
R =\frac{1}{E}\sum_{n=-N}^{-2}e_n, \quad T = \frac{1}{E}\sum_{n=2}^N e_n, \quad
E_\text{tr}=1 - T - R = \frac{1}{E}\sum_{n = -1}^{1} e_n,
\eeq
where we have (admittedly, with some degree of arbitrariness) assumed that the defective region
encompasses the defect site and its nearest neighbors, while energy fractions to the left and
right of that represent, respectively, the reflected and transmitted contributions to the energy.

While the above setting includes a single defect, it is straightforward to generalize it to the
case of multiple resonators, as discussed below. Having provided the theoretical setup, we now turn
to experimental results in a woodpile chain with one and two defects as a motivation for the detailed
theoretical and computational studies that will follow.

\section{Experiments Using Simple 1D Woodpile Lattices }
%

The experimental construction of a relevant setup requires a lattice system that features
(i) nonlinear Hertzian contact among constituent particles and (ii) localized linear
oscillators in defect sites. In a recent study of~\cite{bonanomi} it was attempted to
build such a system using a chain of hollow spheres that embed resonators by using polymeric
holders. While this system demonstrated the feasibility of developing tunable frequency
bandgaps, it is susceptible to damping due to the viscoelastic nature of the polymeric
holders. Another complex system, a bead having an attached ring resonator, has also been
reported in~\cite{gatz}. Some of the present authors built a 1D woodpile lattice system~\cite{eunho01},
which derives the local resonances of constituents from the bending vibrations of longitudinal
woodpile elements. By using this setup, we have successfully verified the versatile propagation of
nonlinear waves, which transmit, modulate, or attenuate depending on the interplay between the
propagating nonlinear waves and the local resonances of woodpile components~\cite{eunho02}. This
also led to the experimental verification of highly nonlinear weakly nonlocal waves (often referred
to as nanoptera) in the setting of homogeneous 1D woodpile lattice systems. The advantage of this
system is that the propagating waves suffer minimal dissipation, so that the dynamics  can be
undisputably verified for this system using non-contact methods, such as laser Doppler vibrometry.
In the present study, we employ such woodpile systems to validate simple representative cases
of the lattice with one or more internal resonator defects, while using numerical simulations
for corresponding parametric studies.

\subsection{Experimental setup}

We built a test setup as shown in Figure~\ref{fig:expSetup} to experimentally demonstrate
the propagation of solitary waves in a woodpile lattice and their interactions with a
defect (i.e., a node bearing a local resonator). In this setup, we use a homogeneous chain
of 40 cylindrical rods including an impurity (or defect) element in the middle of the chain,
i.e., at $n=0$. All rods are made of fused quartz (Young's modulus $E=72$ GPa, Poisson's
ratio $\nu=0.17$, and density $\rho=2,200$ kg/m$^3$) and have an identical diameter of $5$
mm, yielding the contact
coefficient $a=2.47\times 10^{9}\; \text{N/m}^{3/2}$. The length of the regular cylinders is $20$ mm, while the defect rod has a $40$ mm length.
To excite the woodpile lattice, we apply an impact on the top of the chain by dropping a $10$ mm
diameter glass sphere from a $20$ cm drop height (impact velocity is $V=1.98$ m/s).
A soft foam is located at the bottom of the chain to suppress and delay the wave reflection
from the boundary.

To observe the wave propagation in the chain we use a Laser Doppler Vibrometer (LDV) mounted
on an automatic guiding rail. When the striker impacts on the top rod, the piezoelectric sensor
bonded on the surface of the top rod generates voltage, which in turn triggers the LDV and
measures a particle's velocity. We record the particles' velocity profiles one by one in each
impact event and synchronize all collected signals with respect to the trigger signals. This
enables us to visualize the wave propagation in a spatio-temporal map (to be presented and
discussed in the next section).

In this woodpile lattice, the vibrations of cylindrical elements play a role of local resonances,
as noted above. The bending vibration modes are particularly important, since they carry most of
the vibration energy in the frequency domain of our interest (below 50 kHz). The slender
cylindrical rods develop low frequency bending modes, and our previous study showed that they
can be coupled with the propagating nonlinear waves~\cite{eunho02}. The mode coupling mechanism
depends on the resonant frequency of the bending mode, and if the resonant frequency is too high
compared to the characteristic time of the propagating nonlinear waves, the coupling effect
becomes weak. We find that the first bending modes of the 20 mm and 40 mm rods are approximately
55.2 kHz and 15.2 kHz, respectively, based on our previous numerical and experimental investigations~\cite{eunho01,eunho02}.
At an impact excitation, the chain composed of 20 mm rods without defect shows single-side weakly
nonlocal (nanopteronic) solitary waves, which consist of a leading solitary wave and an oscillating
tail behind it~\cite{eunho02}. The mechanical energy contained in this wave tail is negligibly
small compared to that of the solitary wave, thereby we can safely neglect the effect of wave tail
in the present considerations. However, the first bending mode of the 40 mm rod is low enough to be
strongly coupled with the propagating solitary wave. We can take into account its oscillating behavior
for achieving a mass-with-mass effect, and this can be also modeled via a discrete element approach~\cite{eunho01,eunho02}.
In the discrete element model, the effective parameters $M$, $m$, and $K$ for the 40 mm rod are
$0.838$ g, $0.894$ g, and $3,961$ kN/m, respectively, so that the values of the dimensionless
parameters are $\eps \simeq 1.0$ and $\kappa \simeq 0.07$. Here, the primary mass $M$ is close to
the mass of the 20 mm rod ($0.866$ g).

\subsection{Experimental and numerical results for single and double MwM defects}

\begin{figure}
\centering
\includegraphics[scale=0.44]{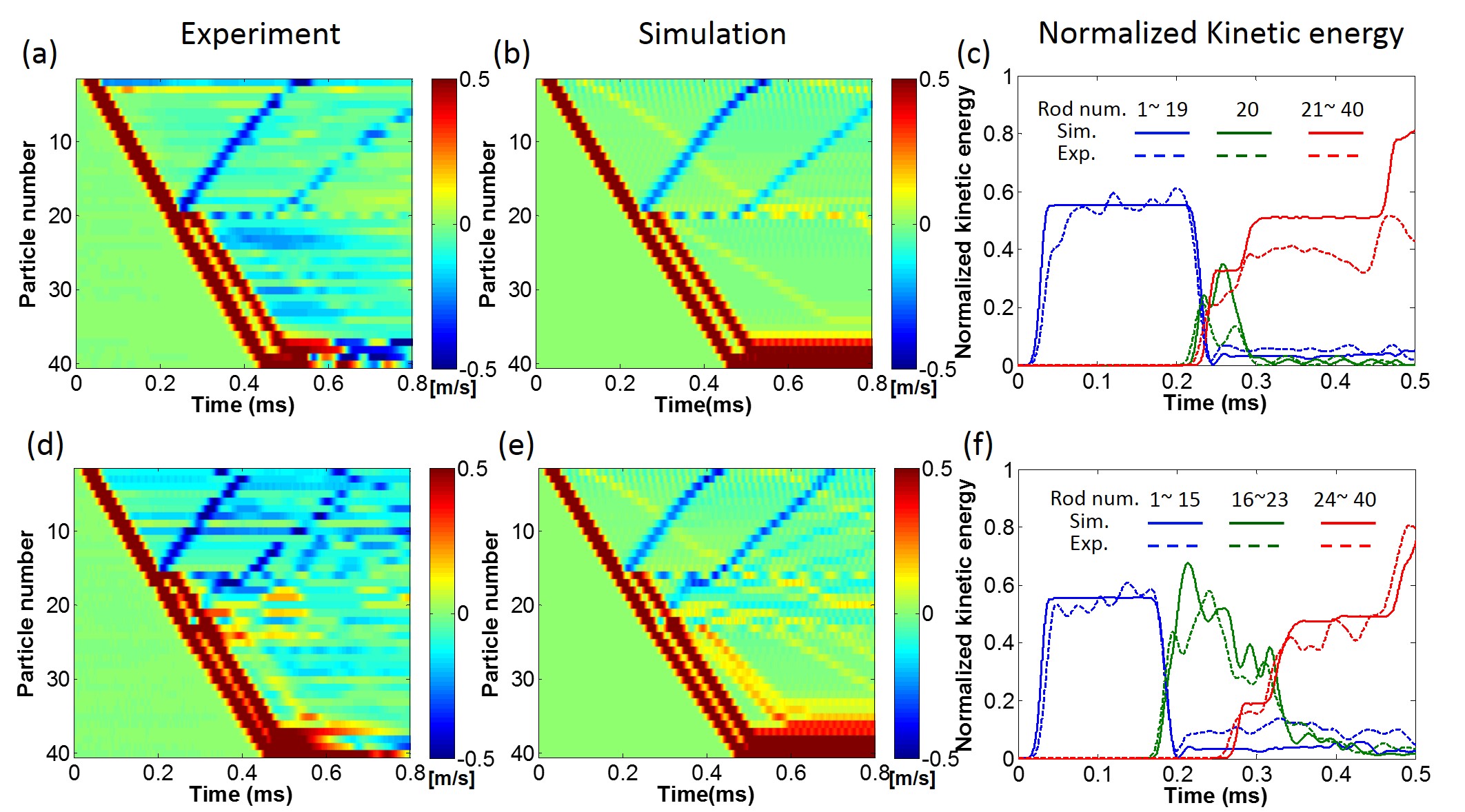}
\caption{(Color online) Solitary wave propagation in the woodpile chain with a single defect (panels (a)-(c)) and two separated defects (panels (d)-(e)) in experiments and numerical simulations. See
the main text for details. Panels (a) and (d) show space-time velocity plots obtained from experiments with one and two defects, respectively,
and panels (b) and (e) show the corresponding results of the numerical simulations.
Panels (c) and (f) compare experimental and numerical evolution of the normalized
kinetic energy fraction in the chain with a single and double defect rods, respectively.}
\label{expfig2}
\end{figure}

We conduct preliminary testing on simple chains containing single and double impurities
bearing local resonators in an otherwise homogeneous chain. Figure~\ref{expfig2}(a) shows
a spatio-temporal map of particles' velocity profiles measured in the chain containing a
single defect rod (40 mm rod) at the center of the chain (20th particle's position from
the top of the chain). The corresponding numerical results for the discrete element model
are shown in Figure \ref{expfig2}(b). Here we use a fourth-order Runge-Kutta method to solve
Eq.~\eqref{eq:orig_eqns} with initial conditions \eqref{eq:orig_ics}, where $V$ is the impact
velocity of the striker and $n_\text{S}=-20$. We use free boundary conditions at both ends
of the chain, which is reasonable in the time span that we are interested in. Note that the
wave reflection at the boundary does not happen in this time frame, as shown in Figure~\ref{expfig2}.

An impact excitation generates multiple solitary waves due to the larger inertia of the striker
compared to the mass of the $20$ mm rod. However, the secondary solitary wave is negligibly small
compared to the primary one, so it is not captured in the experiment. The propagating primary
solitary wave experiences scattering at the defect rod and disintegrates into multiple wave packets.
A part of the energy is reflected back at the defect site due to the larger effective mass of the
defect rod ($40$ mm rod) than that of $20$ mm rod. We also observe that a portion of the energy is
transferred through the defect rod in the form of solitary waves without noticeable time delay.
Interestingly, a fraction of the incident energy is stored in the defect rod, and much of it is released
after a time delay. Therefore a strong secondary solitary wave is generated just behind the primary
transmitted wave. The rest of the energy is trapped in the defect rod in the form of local oscillations,
which slowly disperse to the neighboring rods. Kinetic energy profiles obtained from the experiment
and numerical simulations are presented in Figure~\ref{expfig2}(c), where the energy is normalized
with respect to the impact energy. Here, temporal energy profiles are represented by three different
regions (ahead of the defect rod, at the defect rod, and behind the defect rod), enabling us to quantitatively
compare the energy transmission, reflection, and the energy trapped in the defect rod. In this model,
most kinetic energy is transmitted, while about 0.4\% and 2.0\% of the total kinetic energy are trapped
and reflected at the defect site, respectively; however, it is worthwhile to note that the kinetic energy
in total is not a conserved quantity, hence these fractions are, in principle, time dependent. Here
we only compare kinetic energy instead of total energy because it is difficult to measure potential energy
accurately using the current measurement technique. That is, we measure particles' velocities using LDV,
and the calculation of particles' relative displacements based on velocity data is susceptible to integration
errors. Figures~\ref{expfig2}(d) and~\ref{expfig2}(e) show experimental and numerical profiles of wave propagation,
respectively, when there are two defects (at the 16th and 23rd particle positions from the top of the chain)
with seven particles' distance in the chain. In this model, multiple scattering events appear to arise in
each of the defect rods and more energy is reflected (about 4\%) and trapped between the two defects (about 2\%) in a long time.

\section{Further Numerical Findings}

Motivated by the experimental investigations of the woodpile lattice with one or two
resonator defects we now consider a series of more detailed computational studies.

\subsection{Increasing the Size of the Defect Region}

\begin{figure}
\centering
\includegraphics[height=.20\textheight, angle =0]{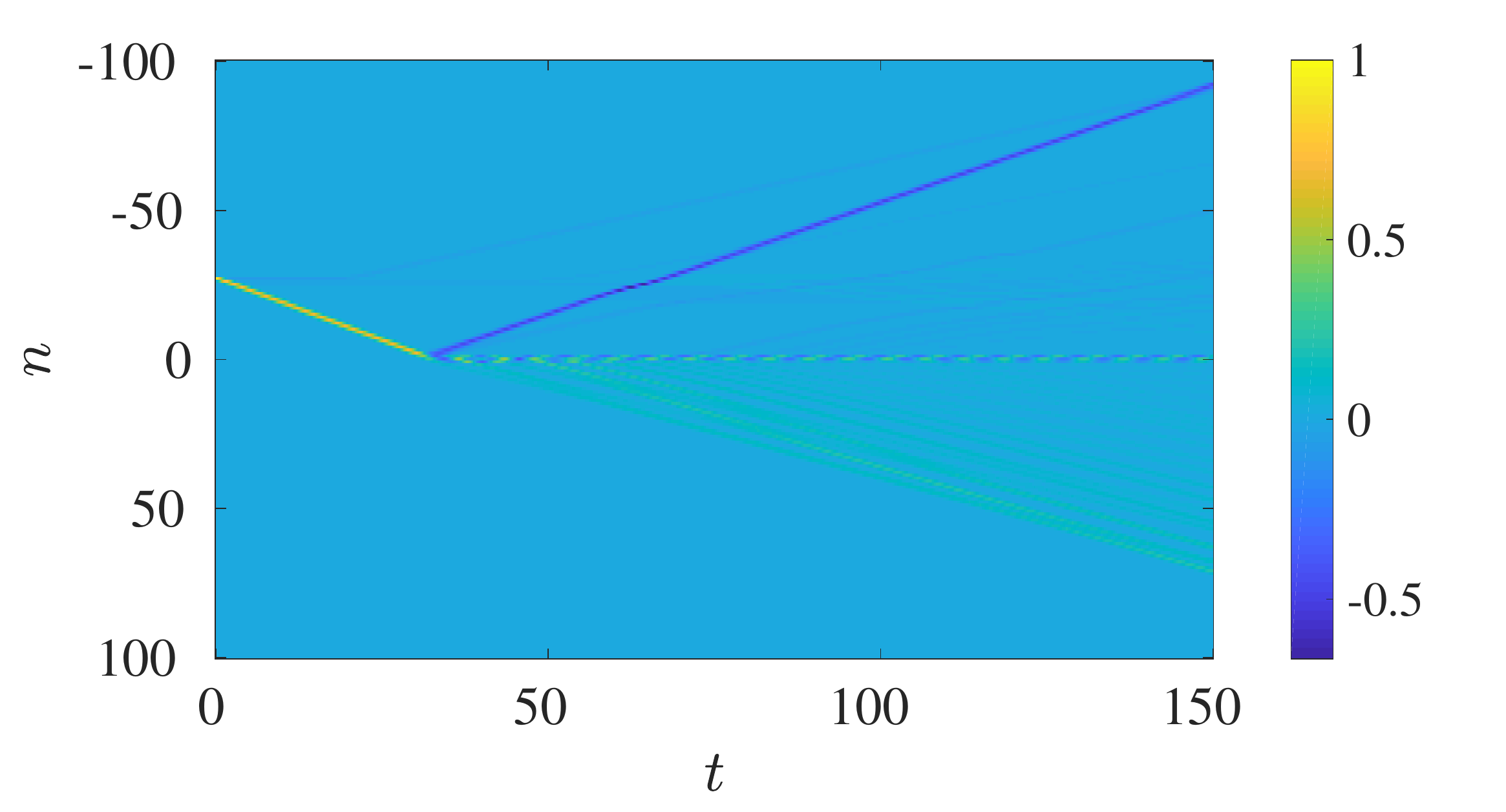}
\includegraphics[height=.20\textheight, angle =0]{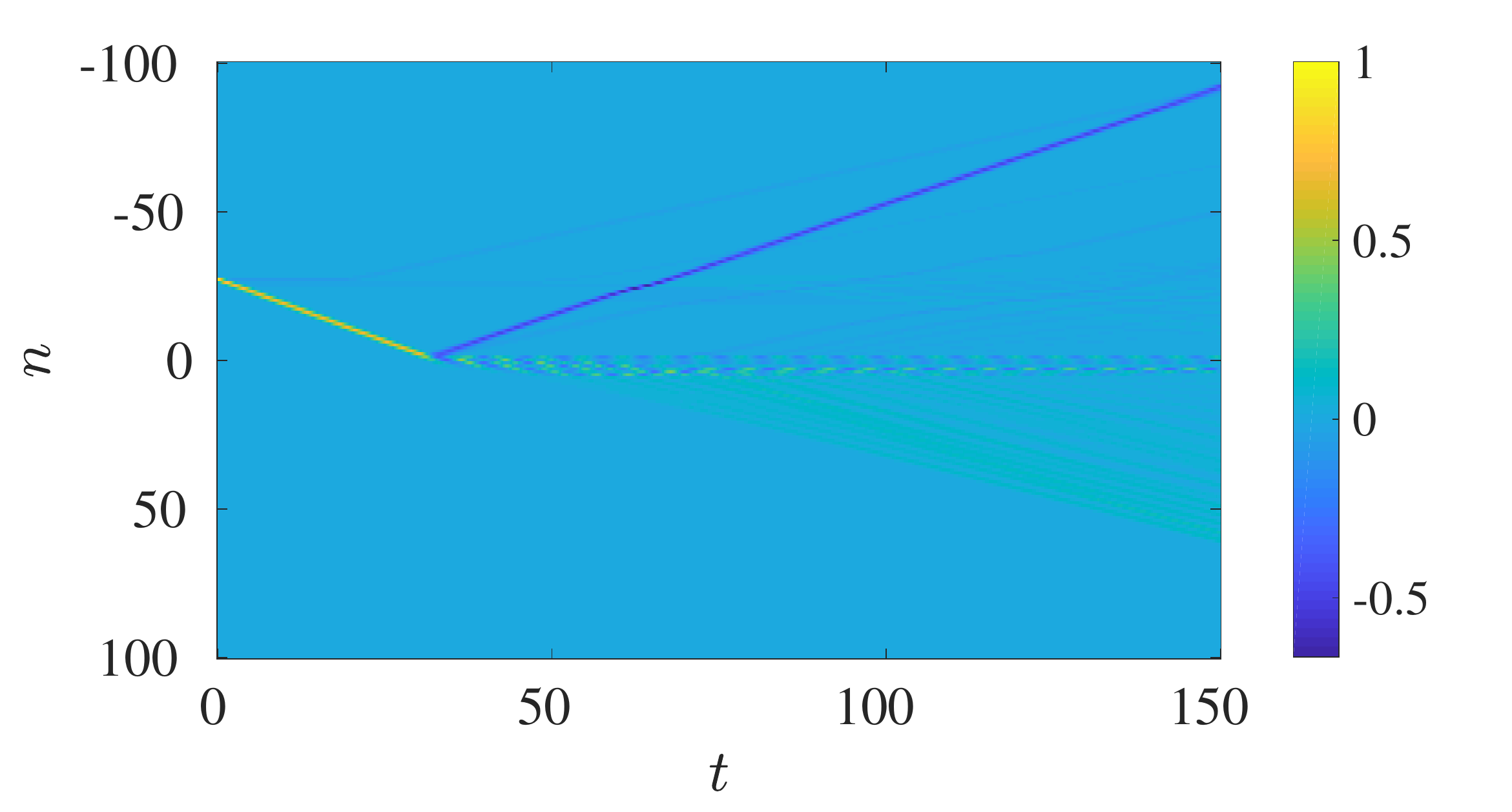}
\caption{(Color online) Space-time evolution of velocities in numerical simulations with $\epsilon = 10$, $\kappa=1$ and defect lengths
$3$ ($L=2$, left panel) and $7$ ($L=6$, right panel). In the second case the two edge defects are visibly more separated
demonstrating a wider defect zone.
\label{numfig2}
}
\end{figure}

We first look at a more systematic exploration of the possibility of having more than
one defects of this resonator type. In this context, we monitor how the presence of
additional defects affects the reflected, transmitted, and trapped fractions of the
energy. Notice that, contrary to the experimental setup, which can more accurately capture
the velocities and hence the kinetic energy, here we use as a more adequate diagnostic
(due to its total conservation) the full energy of the system. The present scenario is modeled by
\beq
\begin{split}
\ddot{u}_n &= \left[u_{n-1} - u_n\right]_+^{3/2} -%
\left[u_{n} - u_{n+1}\right]_+^{3/2} - \sum_{j=0}^{L}\kappa (u_j - v_j)\delta_{nj}, \\
\epsilon\ddot{v}_j &= \kappa (u_j - v_j), \quad j=0,\dots, L,
\end{split}
\label{eqmotion}
\eeq
where $L+1$ is the number of defects. Using a fourth-order Runge-Kutta method, we numerically
integrated Eq.~(\ref{eqmotion}) forward in time with $N=100$ and $0 \leq L \leq 9$, with zero
boundary conditions and zero initial displacements and velocities except $\dot{u}_{-25}(0)=1$.
Figure~\ref{numfig2} displays the space-time evolution of velocities for $\epsilon = 10$
and $\kappa = 1$, with left and right panels corresponding to $L=2$ and $L=6$, respectively.

\begin{figure}
\begin{center}
\mbox{\hspace{-0.7cm}
\subfigure[][]{\hspace{-0.2cm}
\includegraphics[height=.18\textheight, angle =0]{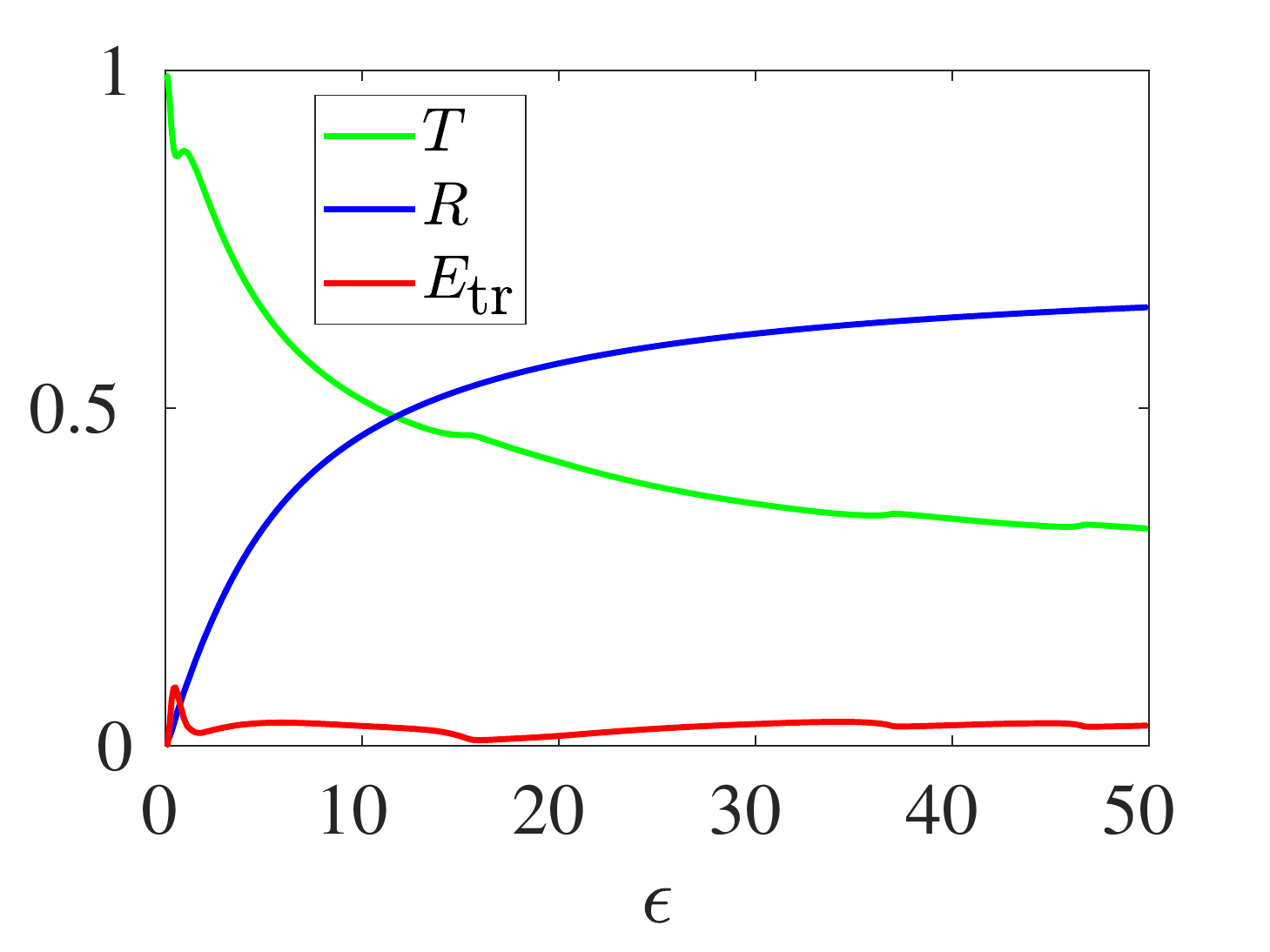}
\label{fig4a}
}
\subfigure[][]{\hspace{-0.5cm}
\includegraphics[height=.18\textheight, angle =0]{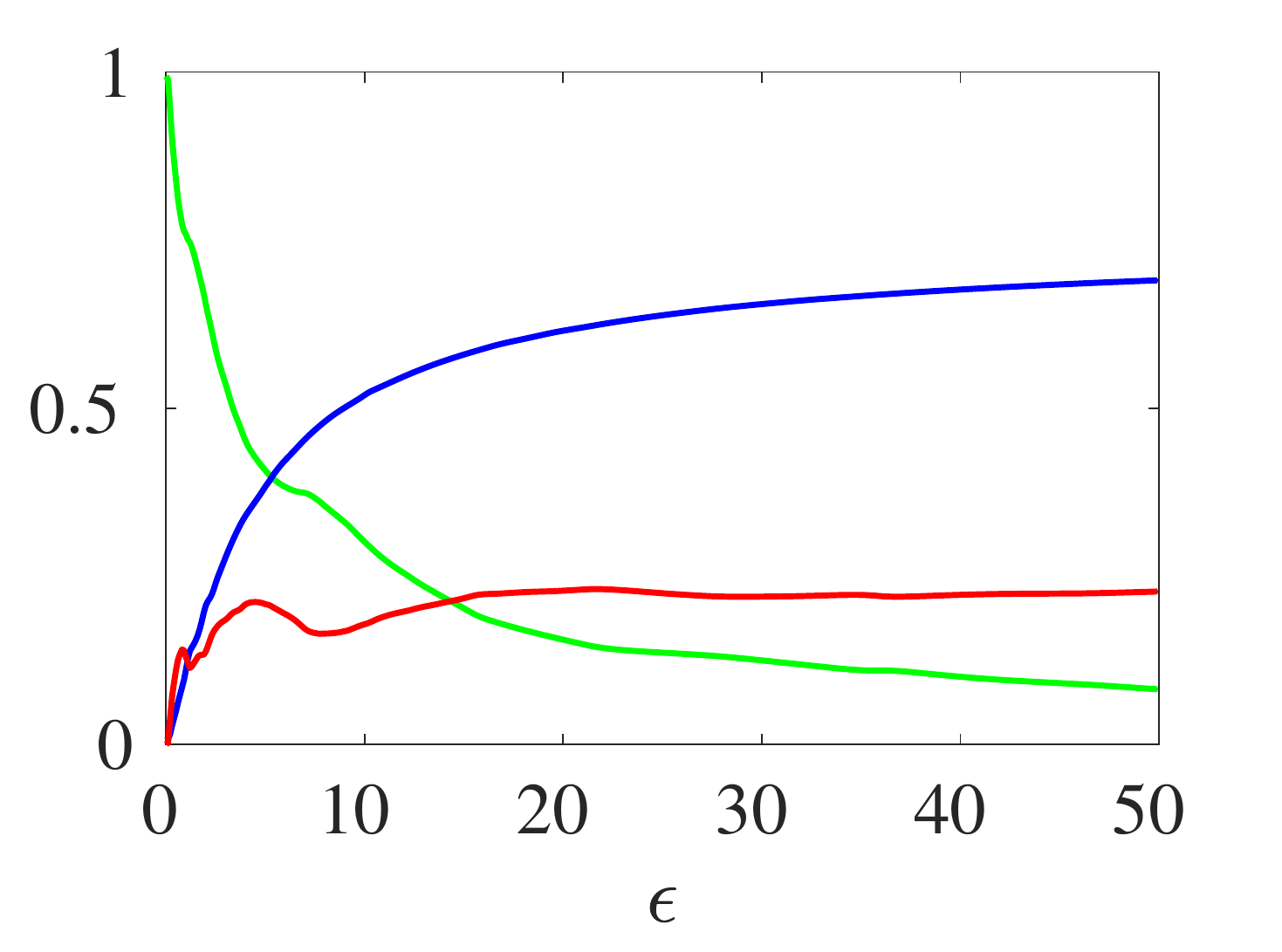}
\label{fig4b}
}
}
\mbox{\hspace{-0.7cm}
\subfigure[][]{\hspace{-0.2cm}
\includegraphics[height=.18\textheight, angle =0]{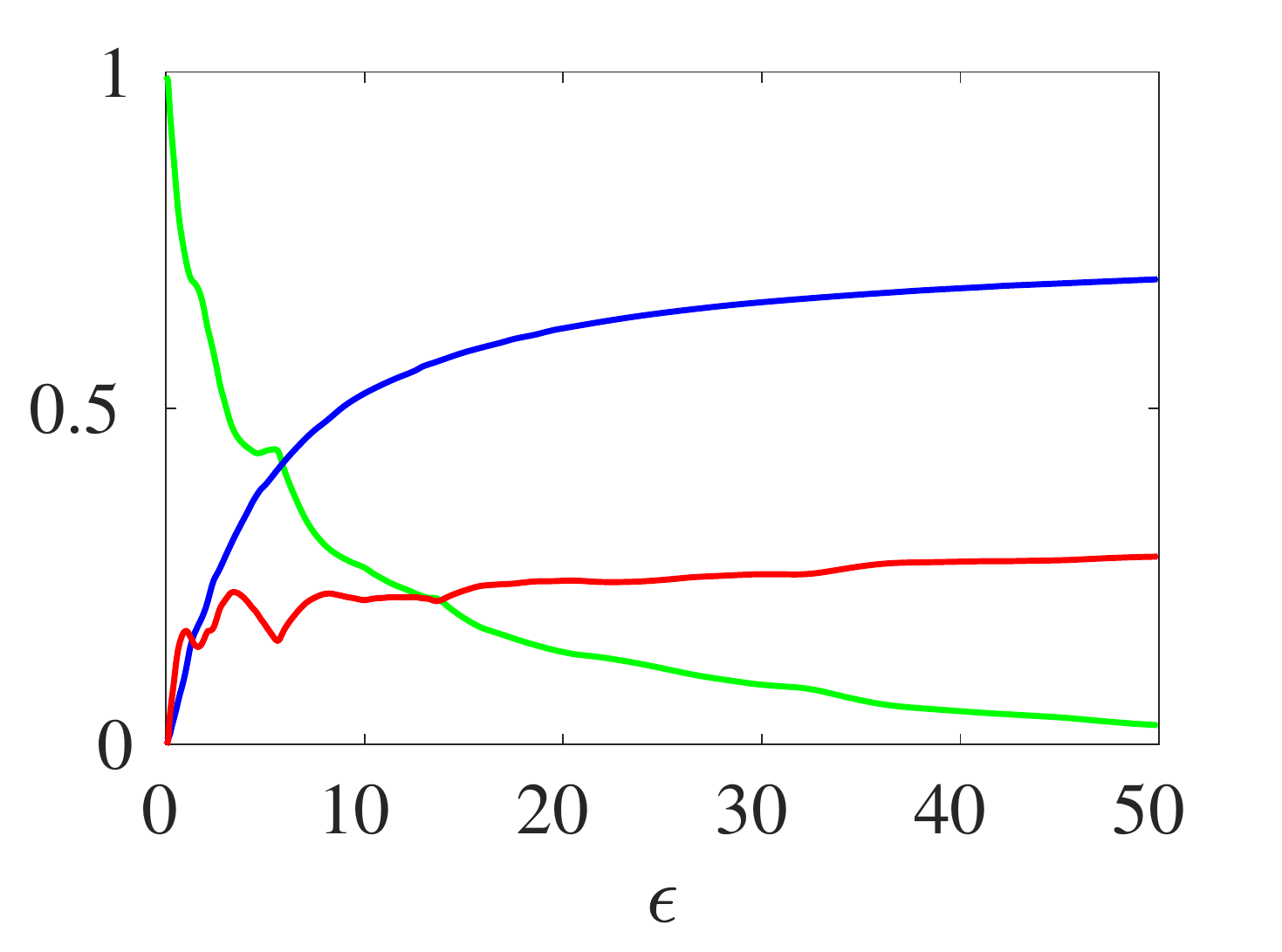}
\label{fig4c}
}
\subfigure[][]{\hspace{-0.5cm}
\includegraphics[height=.18\textheight, angle =0]{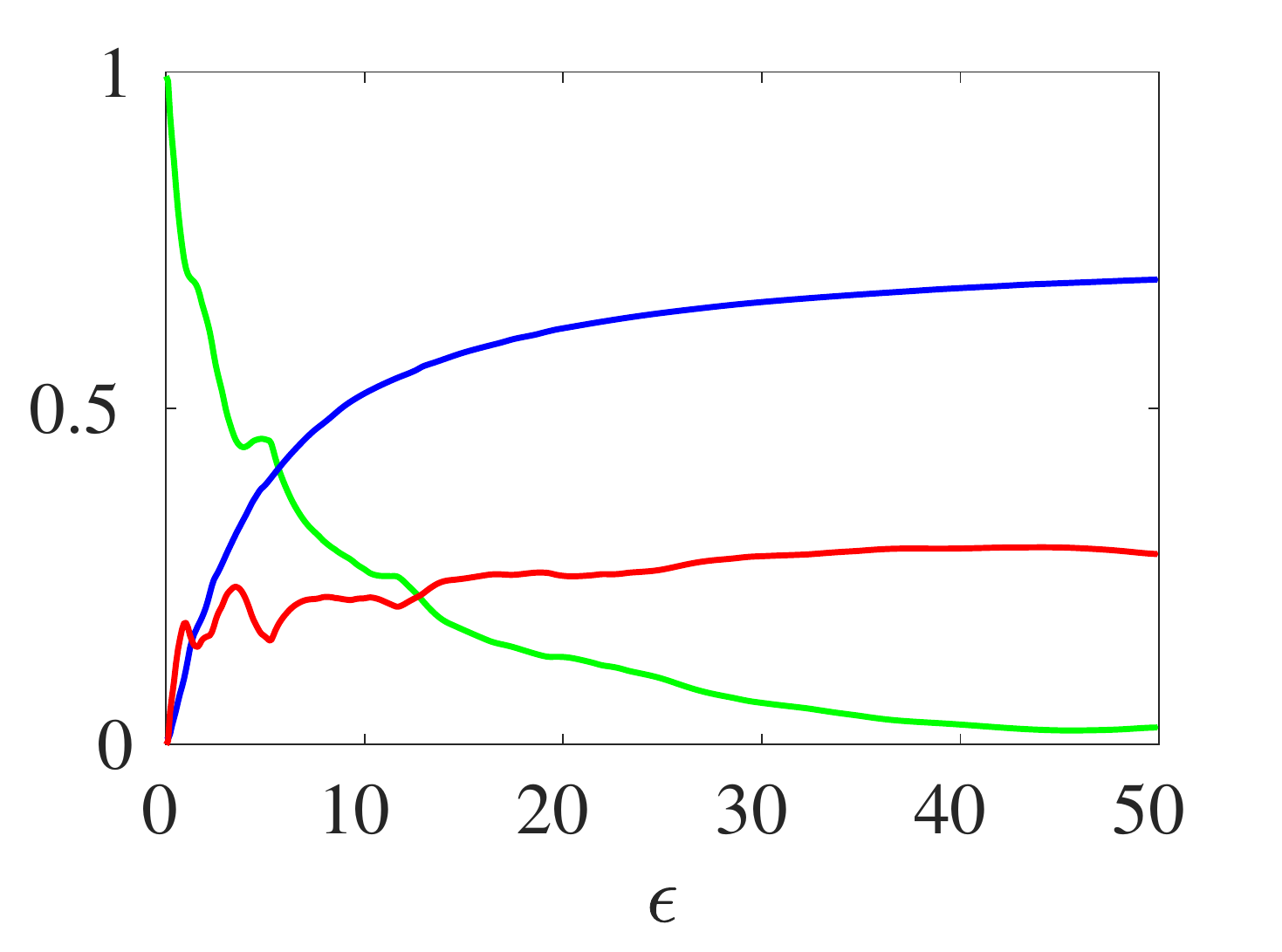}
\label{fig4d}
}
}
\end{center}
\caption{(Color online) The reflected ($R$), transmitted ($T$) and
trapped ($E_{\textrm{tr}}$) energy fractions as functions of the mass $\epsilon$
with $\kappa=1$  for the cases of: (a) single defect ($L=0$), (b) $3$
adjacent defects ($L=2$), (c) $6$ adjacent defects ($L=5$), and (d) $7$
adjacent defects ($L=6$).
\label{numfig3}
}
\end{figure}

Figure~\ref{numfig3} shows how the reflected, transmitted and trapped fractions
of the energy, given by
\beq
R=\frac{1}{E}\sum_{n=-N}^{-2} e_n, \quad T=\frac{1}{E}\sum_{n=L+2}^{N} e_n
\eeq
and $E_\text{tr}=1-T-R$, respectively, and
\begin{equation}
e_n =\frac{1}{2}\dot{u}_n^2 + \frac{1}{2}\sum_{j=0}^{L}\left[\eps\dot{v}_j^2 %
+ \kappa (u_j - v_j)^2\right]\delta_{nj} +%
\frac{1}{5}\Big\lbrace\left[u_{n-1}-u_n\right]_+^{5/2}+\left[u_n - u_{n+1}\right]_+^{5/2}\Big\rbrace,
\end{equation}
vary as functions of the mass of the defect $\epsilon$ for the cases of $1$, $3$, $6$,
and $7$ resonators composing the relevant ``defective region'' within the chain for
$\kappa = 1$. To obtain the parametric variation results shown in the figure, we increase
$\epsilon$ by $\Delta \epsilon=0.1$ and evaluate the energy fractions long after the interaction 
of the incoming wave with the defect (i.e., the time integration is performed for $t\in [0,150]$).

\begin{figure}
\centering
\includegraphics[height=.18\textheight, angle =0]{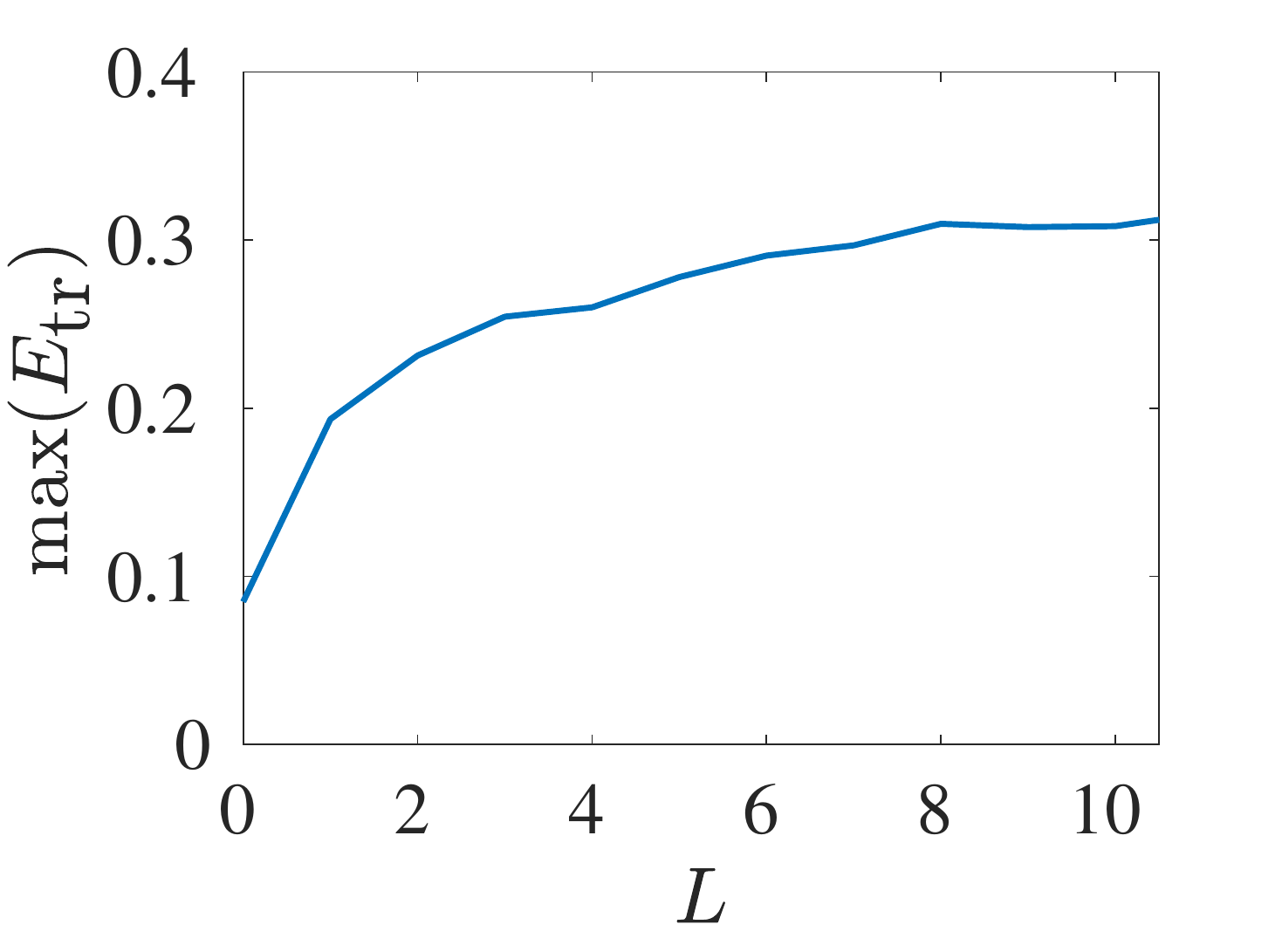}
\caption{(Color online) The global maximum (over the mass of the
resonator $\epsilon$) of the trapped energy as a function of the
length $L$ of the ``defective'' region (bearing the resonators),
with $\kappa = 1$.
}
\label{numfig4}
\end{figure}

In all panels, as $\epsilon\rightarrow 0$, physically corresponding to the case where the resonators
are essentially absent from the system,  we have almost perfect transmission of the energy which is
observed for all $L$. However, the cases with multiple adjacent defects are different from the single-defect
case for large values of $\epsilon$. Generally, when $\eps \gg 1$, there is almost no transmitted energy,
while there is a large amount of reflected energy. Importantly for our considerations involving the
question of how much energy can be trapped in the resonator region, we see a significant increase (as
well as, arguably, a more complicated functional dependence) of this fraction on $\epsilon$ as the number
of resonators increases. We examine as diagnostics both the global maximum of this trapped fraction, as
well as the $\epsilon$ for which it occurs. This is shown in detail in Figure~\ref{numfig4}, presenting 
the relevant (global) maximum for $\epsilon\in(0,60]$. Despite the somewhat non-smooth nature of the relevant 
graph, overall the trend is apparent and illustrates a concave down dependence of the associated fraction of 
the energy on the length of the relevant region within the chain.

\subsection{Increasing the Power of the Resonator-Lattice Interaction}

We now consider how the results are modified if the resonator present in the chain is not
a linear one as assumed up to this point, but rather involves a nonlinear coupling, with
interaction power $p$. In this case, the model equations of interest become
\[
\begin{split}
\ddot{u}_n &= \left[u_{n-1} - u_n\right]_+^{3/2} - %
\left[u_{n} - u_{n+1}\right]_+^{3/2} - \kappa (u_0 - v_0)^p\delta_{n0}, \\
\epsilon\ddot{v}_0 &= \kappa (u_0 - v_0)^p,
\end{split}
\]
where the energy density is given by
\begin{equation}
e_n = \frac{1}{2}\dot{u}_n^2 + \frac{\epsilon}{2}\dot{v}_0^2\delta_{n0} + %
\frac{\kappa}{p+1}(u_0 - v_0)^{p+1}\delta_{n0} %
+\frac{1}{5}\Big\lbrace[u_{n-1}-u_n]_+^{5/2}+[u_n - u_{n+1}]_+^{5/2}\Big\rbrace.
\end{equation}

\begin{figure}
\centering
\includegraphics[height=.18\textheight, angle=0]{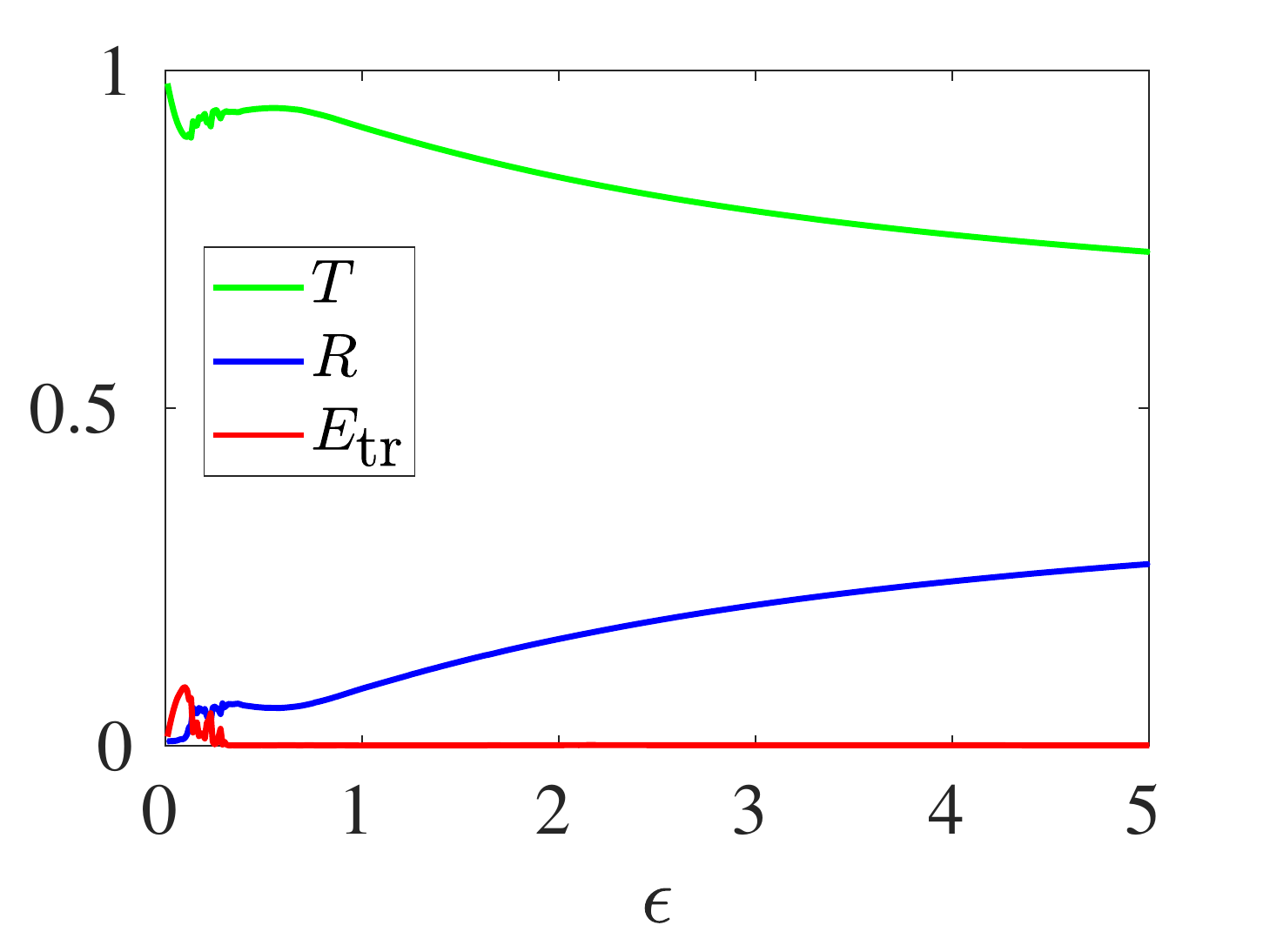}
\includegraphics[height=.18\textheight, angle=0]{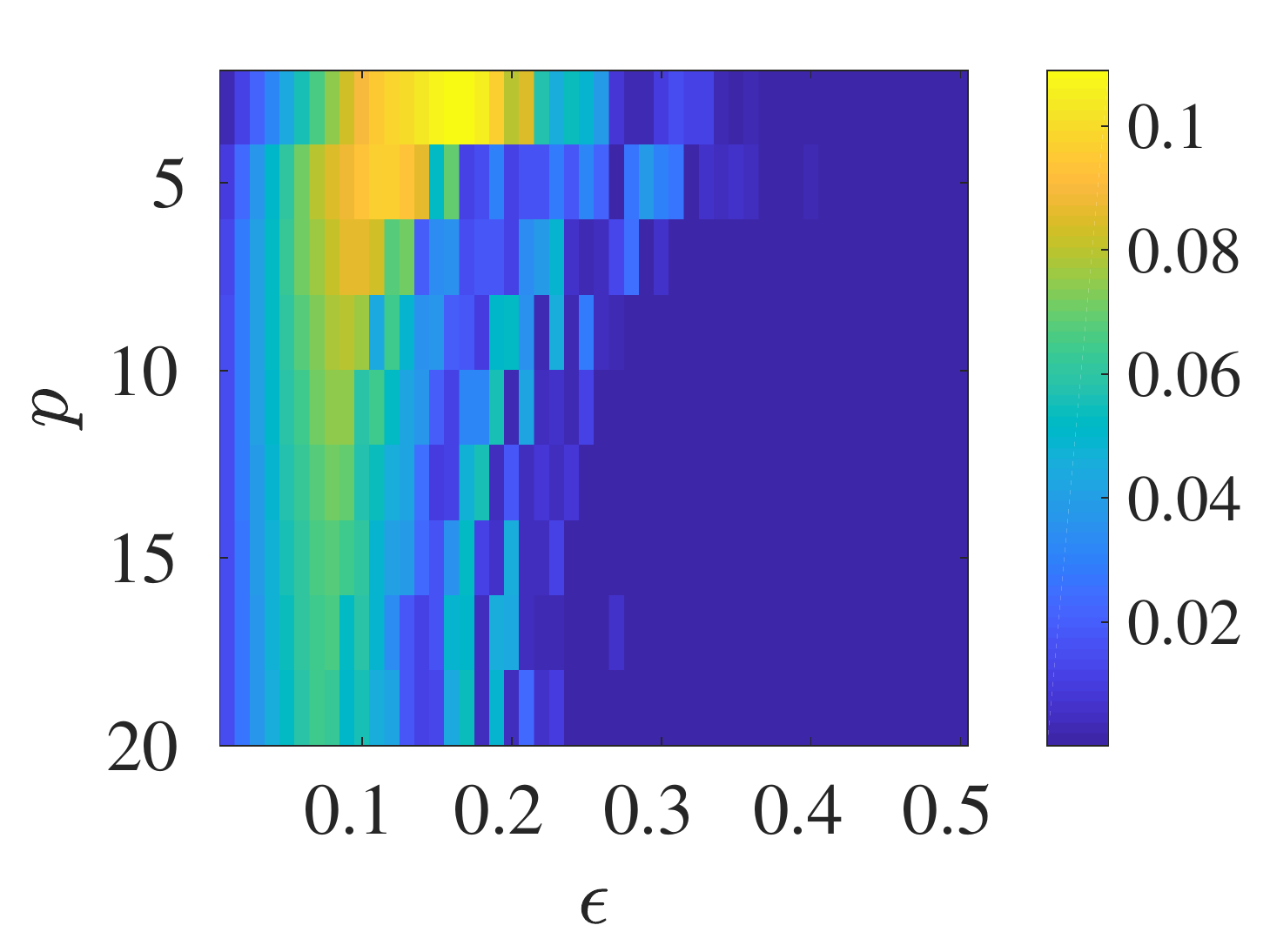}
\caption{(Color online) The left panel shows the reflected, transmitted,
and trapped fractions of the energy for $p=7$, while the right panel
yields a contour plot of the trapped energy for $\epsilon\in(0,0.5]$,
with a fine step of $\Delta\eps=0.01$, and odd $p$ values ranging from
$3$ to $19$. Here $\kappa=1$.
\label{numfig5}
}
\end{figure}

To preserve the restoring character
of the relevant force, we restrict our considerations to
the case of odd integer $p$ ranging from $3$ to $19$.
Other dynamical run parameters
are the same as in the earlier numerical experiments. Figure~\ref{numfig5} shows the different fractions of
the energy for $p=7$, while the right panel of the figure shows how the trapped energy varies in
the two-parameter space of $(\epsilon,p)$. There, it is worthwhile to note that the minimum occurs quite
close to $\epsilon=0$; notice the apparent yellow contour over
the two-dimensional parametric space.
As $p$ increases, the
trapped fraction of the relevant energy decreases for a given
$\epsilon$.
This may be natural to expect in this case on the basis of the
increased ``stiffness'' of the oscillator.
The simulation results shown in Figure~\ref{numfig5}
suggest that in the associated parameter region the optimum values maximizing the amount of trapped energy
are given by $p=3$ and $\epsilon=0.16$ and yield $E_{\textrm{tr}} \approx 0.1088$.

\section{Separating the Defects: Forming an Energy-Trapping Region}

Going back to the linear coupling setting, we now consider another important case when
two defects are placed at a certain number of beads apart. This represents a generalization of the problem with
adjacent defects discussed above. The goal in this case is to be able to trap, to the degree
possible, the energy of a propagating wave between the two defect sites, $n=d_1$
and $n=d_2$, at which the primary unit (upon rescaling) masses are coupled to secondary masses
$\eps_1$ and $\eps_2$ with displacements $v_{d_1}$ and $v_{d_2}$, respectively, and, for simplicity, 
same coupling parameter $\kappa$. Our trapped energy
($E_\text{tr}=1-T-R$) maximization then leads us to seek maximizing the quantity
\begin{eqnarray}
E_\text{tr} = \frac 1 E \sum_{n = d_1 -1}^{d_2+1} e_n,
\label{maximize}
\end{eqnarray}
where the energy density $e_{n}$ in this case is given by
\begin{equation}
e_n =\frac{1}{2}\dot{u}_n^2 + \frac{1}{2}\sum_{j=1}^2\left[\eps_j\dot{v}_{d_j}^2 +%
\kappa (u_{d_j} - v_{d_j})^2\right]\delta_{nd_j}+%
\frac{1}{5}\Big\lbrace\left[u_{n-1}-u_n\right]_+^{5/2}+\left[u_n - u_{n+1}\right]_+^{5/2}\Big\rbrace,
\end{equation}
and $E$ is the total energy.

\begin{figure}
\centering
\includegraphics[height=.18\textheight, angle=0]{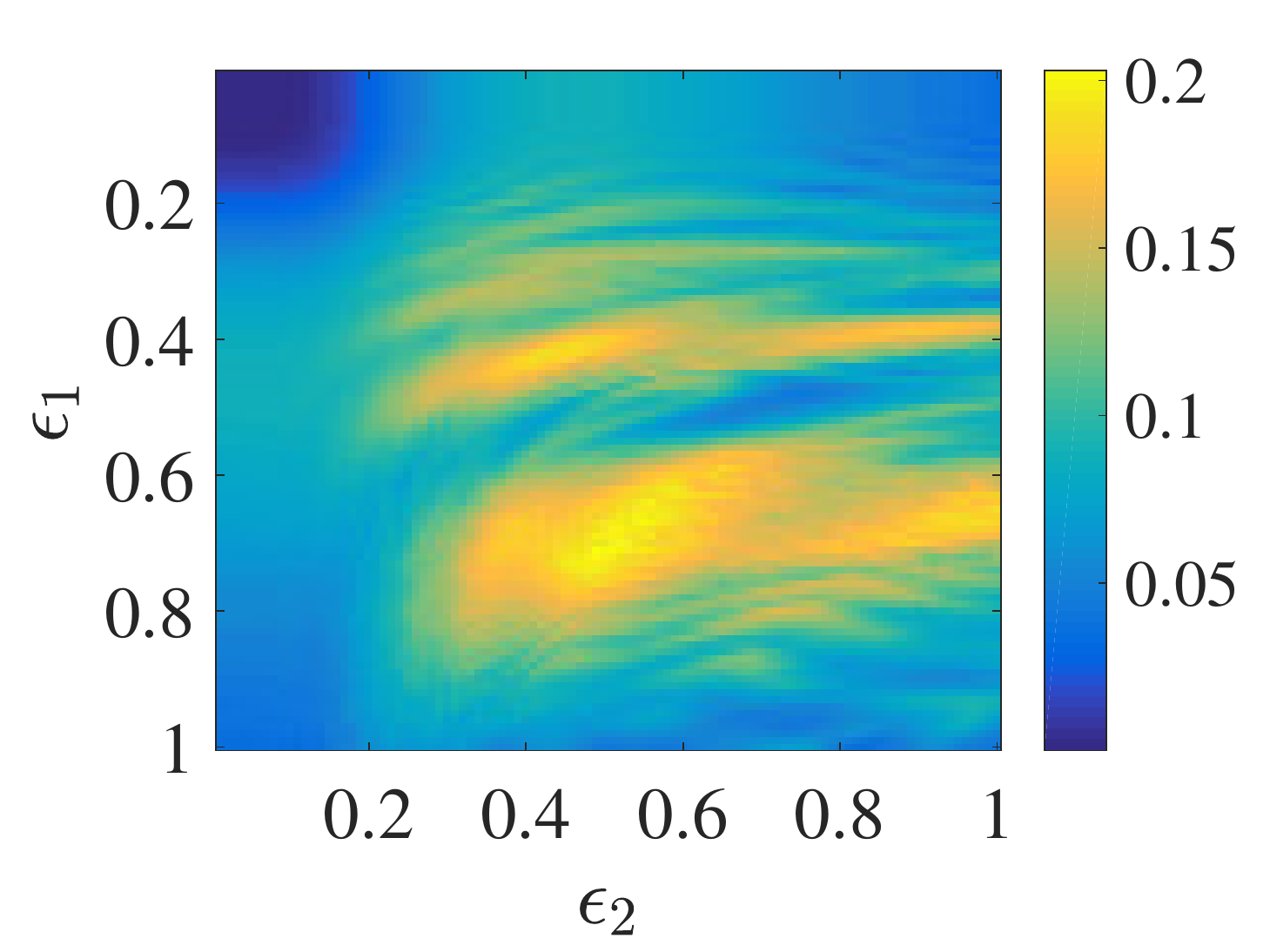}	
\includegraphics[height=.18\textheight, angle=0]{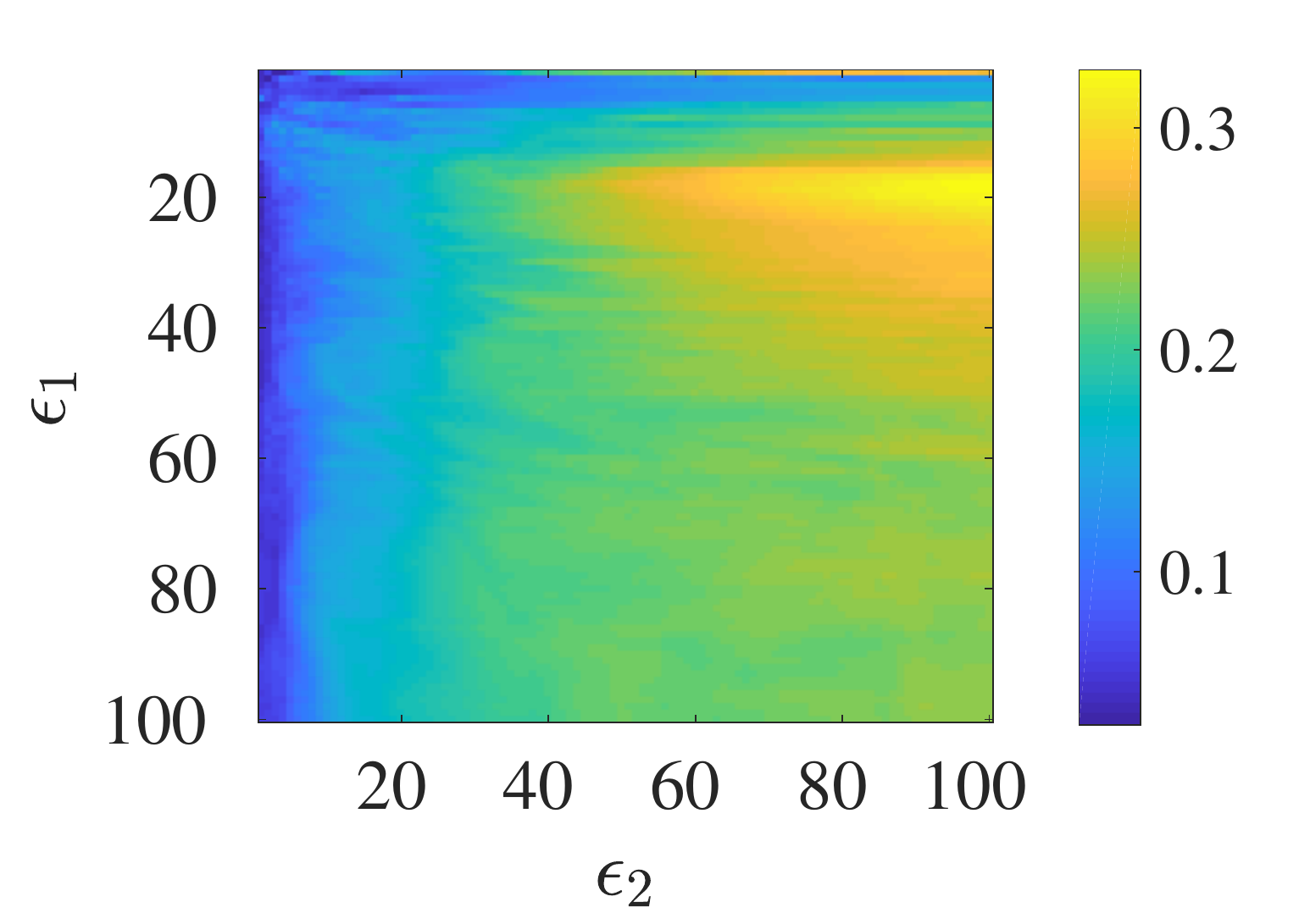}
\caption{(Color online) Contour plot of the trapped energy for various $\epsilon_1$
and $\epsilon_2$ ratios, as a two-dimensional function of both variables (for fixed
$\kappa = 1$). The left panel is for a finer grid spacing and smaller yet important
(in terms of energy trapping) region of the two-dimensional parameter space, while
the right panel represents a coarser but extended grid. The left panel shows the domain
$[0.1,1]^{2}$ with an $\epsilon$ increment of $\Delta\epsilon=0.01$ and the right panel 
shows the domain $[1,100]^{2}$ with an $\Delta\epsilon=1$.
\label{numfig7}
}
\end{figure}

\begin{figure}
\centering
\includegraphics[height=.20\textheight, angle=0]{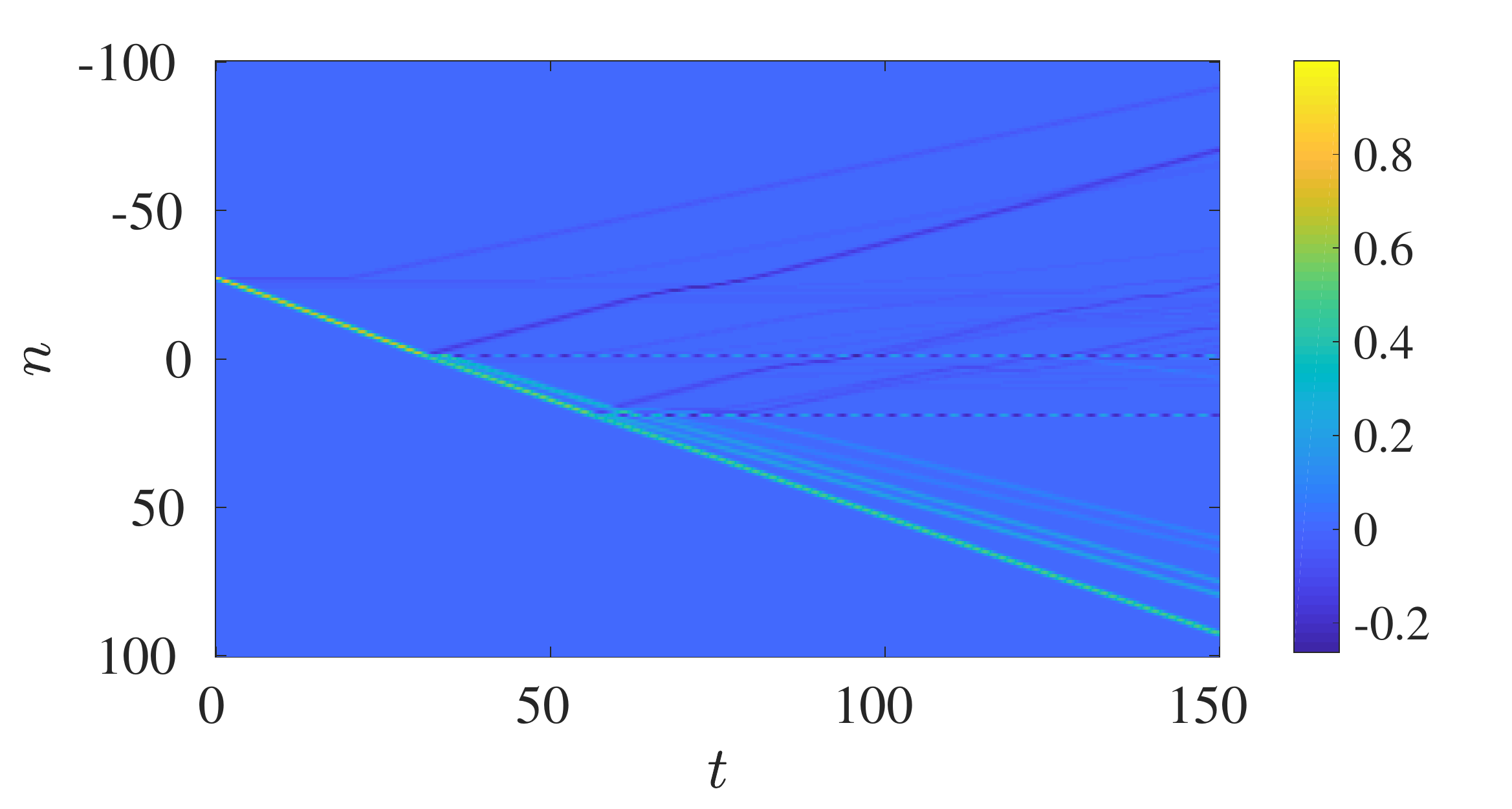}
\includegraphics[height=.20\textheight, angle=0]{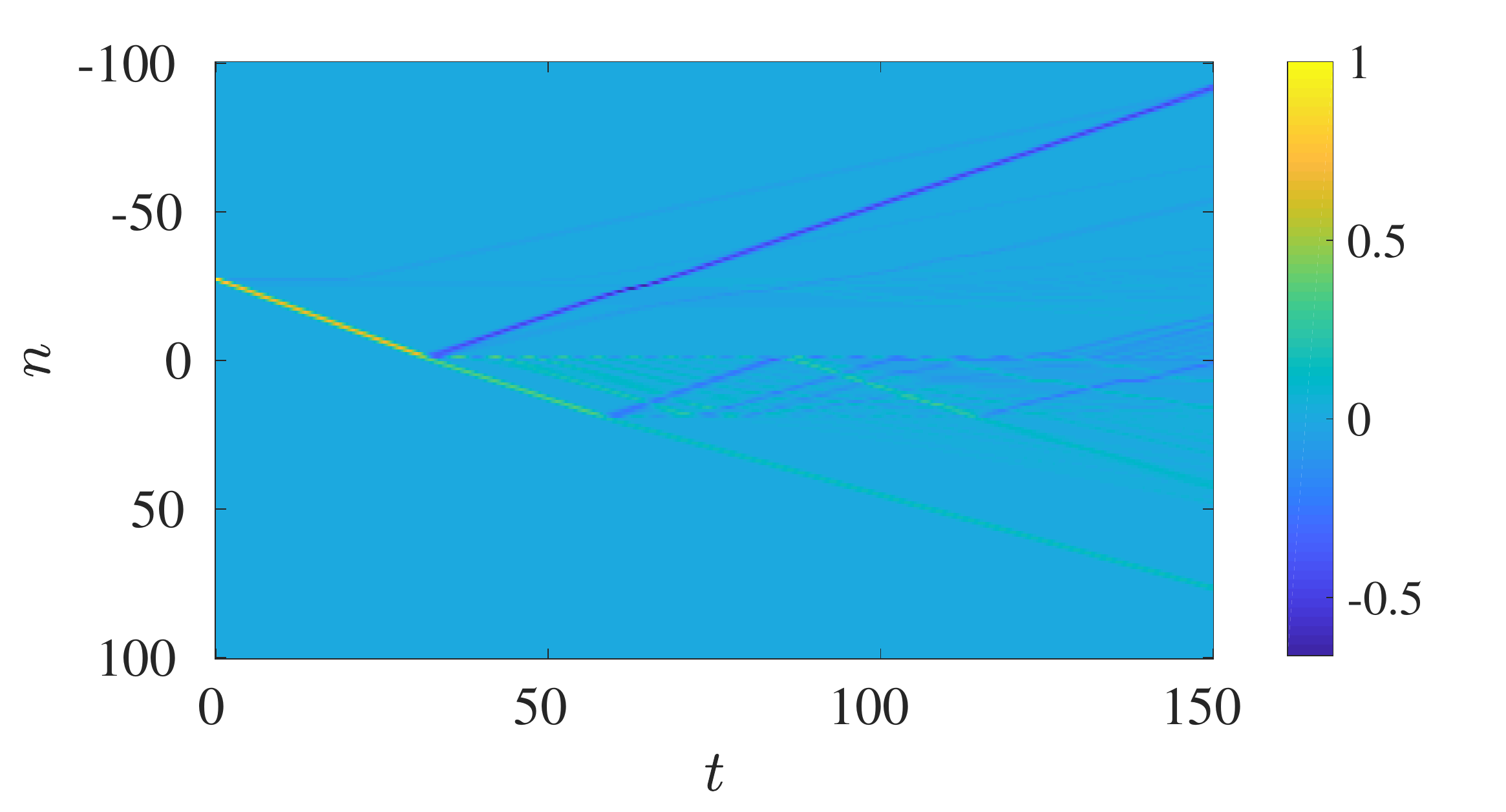}
\caption{(Color online) Space-time velocity plot
for the $(\epsilon_1, \epsilon_2)$ combinations that
yield the maximal trapping. The left panel corresponds to the
(ordered) pair $(\epsilon_1,\epsilon_{2}) = (0.67,0.54)$ while
the right panel to $(\epsilon_1,\epsilon_{2}) = (13,100)$. In
the left panel we see that the major contributor to the trapped
energy is the energy emanating from the resonator nodes, as opposed
to the right panel where we see that the major contributor to the
trapped energy is the motion of a single solitary wave bouncing
back and forth between the beads. In addition, in the right panel
we can also discern slightly smaller solitary waves emanating from
the ringing vibrations of the two defects, as indicated by the darker 
yellow lines.
\label{numfig8}
}
\end{figure}

As a representative example, we consider the wide region between the two defects,
with $d_1 = 0$, and $d_2 = 20$. Running the same simulations as above, we vary the
masses $\epsilon_{1}$ and $\epsilon_{2}$ of the defects and measure the trapped
energy as shown in Figure~\ref{numfig7}. The left panel of the figure corresponds to
a smaller mass ratio domain, $\epsilon_{1}\times\epsilon_{2}=[0.1,1]^{2}$, whereas
the right panel to a larger domain $\epsilon_{1}\times\epsilon_{2}=[1, 100]^{2}$.
For the smaller domain we increment $\epsilon_{1,2}$ by $\Delta\epsilon=0.01$, and for the larger
domain we have $\Delta\epsilon=1$. The maximum of the trapped energy fraction for the
larger domain with the coarser grid (see the figure caption for details)
arises when $\epsilon_1 = 13$ and $\epsilon_2 = 100$ with $E_\text{tr} = 0.3926$, i.e., a
significant component of the energy being trapped in the relevant region.
Importantly, note that in this case the optimum arises on the ``boundary''
of the parametric domain associated with the maximal mass of the second
defect, enabling (presumably) the largest possible (inward) reflection
from that boundary. In the case of the smaller domain with finer mesh it occurs when 
$\epsilon_1 = 0.67$ and $\epsilon_2 = 0.54$ with $E_\text{tr} = 0.2112$.
These cases are illustrated in Figure~\ref{numfig8}. On the finer mass ratio scale
(left panel), the main contribution to the trapped energy stems from the pair of
defect beads which interact  with their respective resonator masses in such a way
as to contain a substantial fraction of the solitary wave as it crosses the defect.
This is somewhat in contrast to the coarser $\epsilon$ case, in which the primary
contributor to the trapped energy is the solitary wave itself, bouncing between the
two defects and resulting in a significant fraction of the energy being confined in
the region between the defects.


\begin{figure}
\centering
\includegraphics[height=.18\textheight, angle=0]{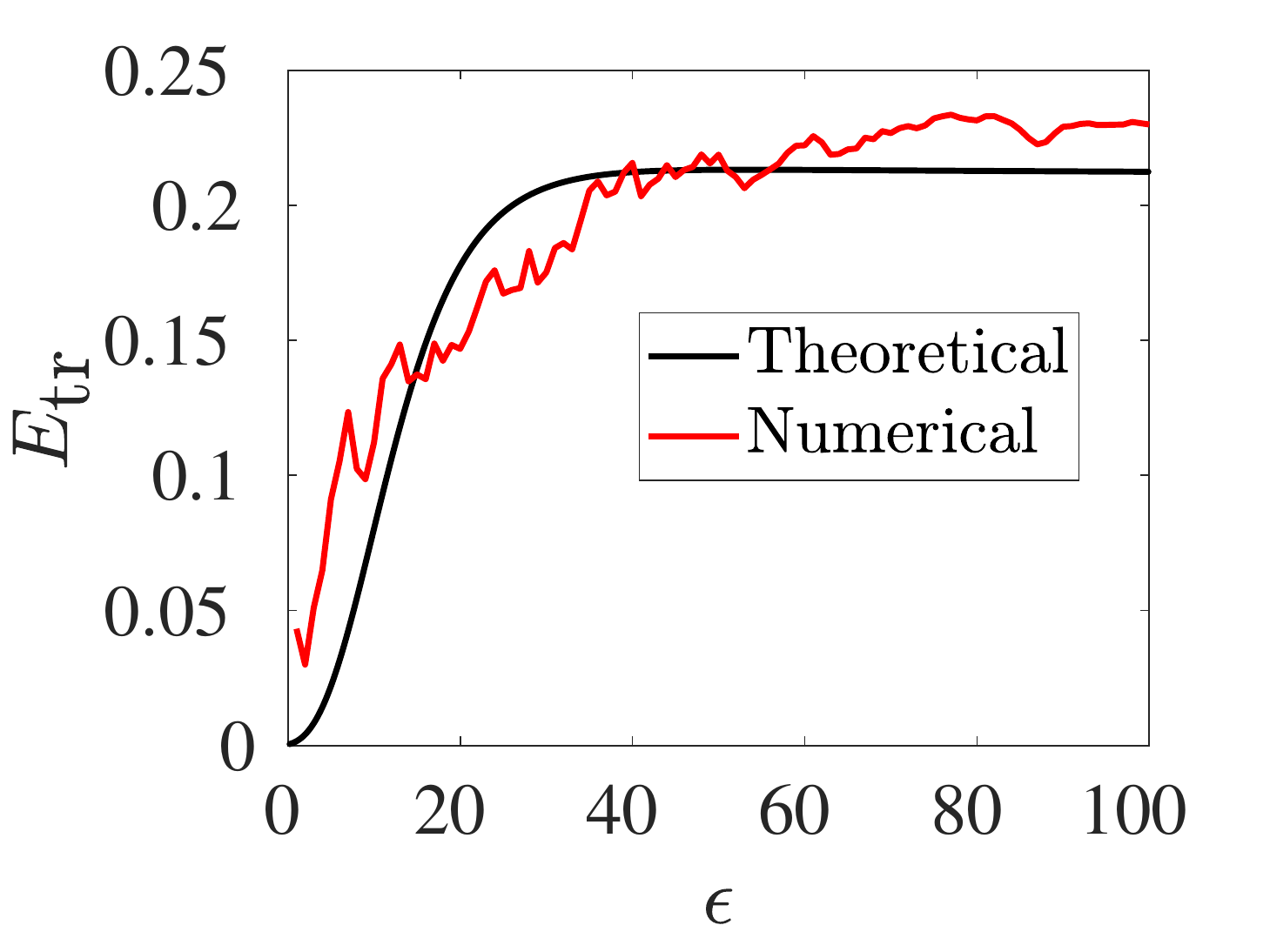}
\includegraphics[height=.18\textheight, angle=0]{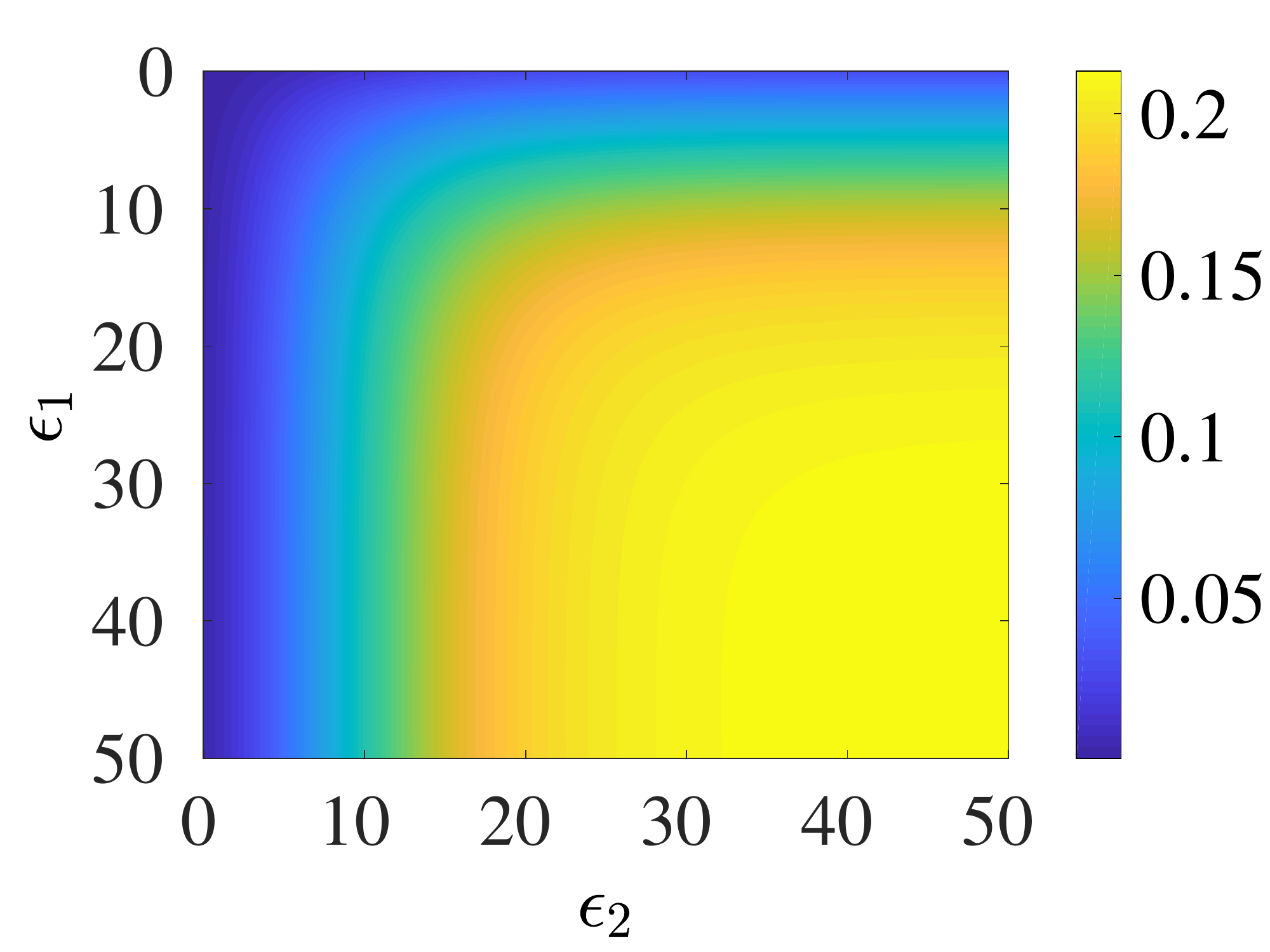}
\includegraphics[height=.18\textheight, angle=0]{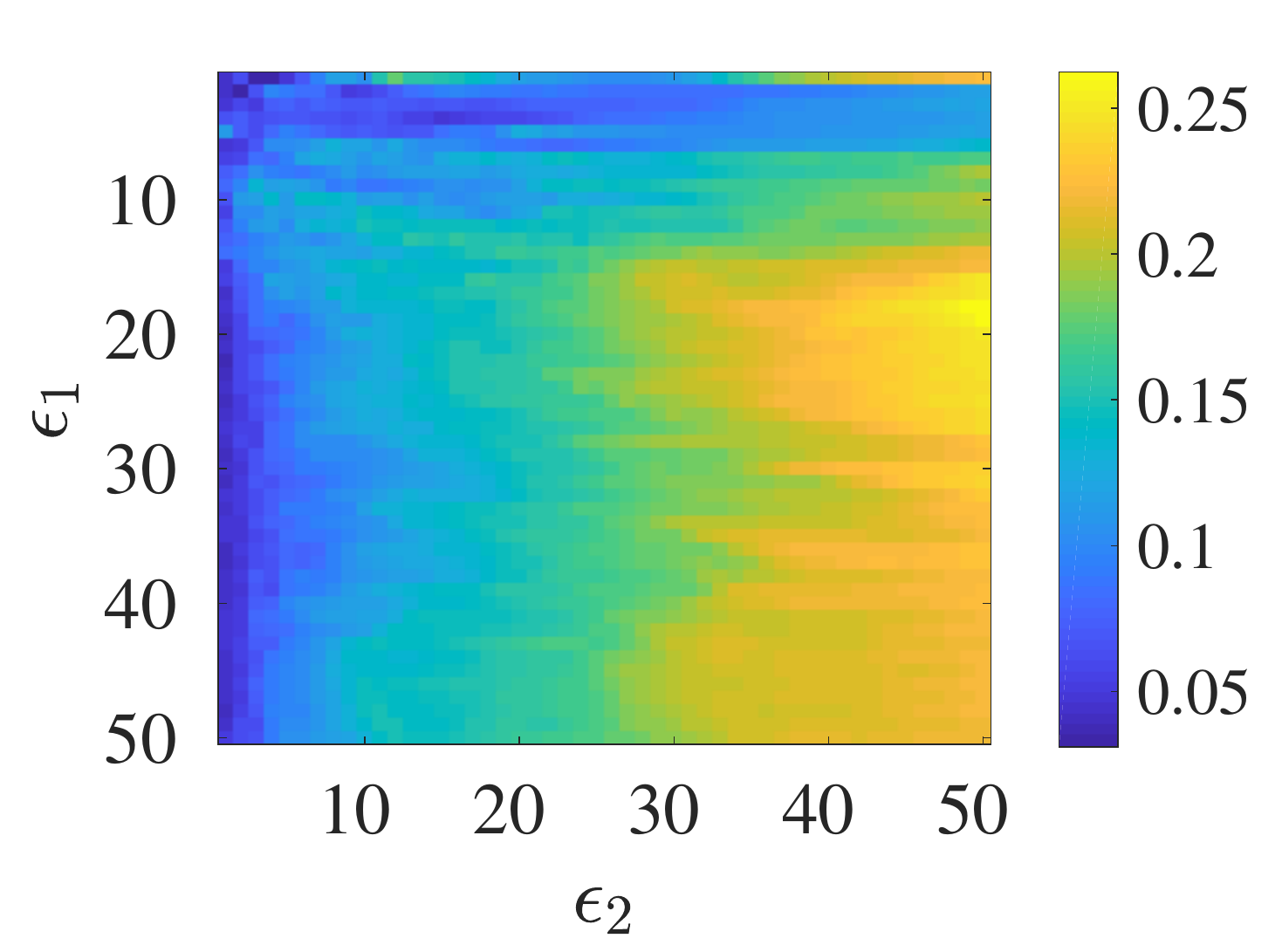}
\caption{(Color online) Left panel: plot of the semi-analytical approximation of $E_\text{tr}(\epsilon)$
discussed in the context of Figure~\ref{numfig7} for $\epsilon_1=\epsilon_2 = \epsilon$.
Middle and right panels: plot of $E_\text{tr}(\epsilon_1, \epsilon_2)$ compared with a zoomed in panel
of Figure~\ref{numfig7}. The middle panel is the analytical approximation while the right panel
represents the numerics.
\label{numfig9}
}
\end{figure}

Returning to our main goal, we would like to
attempt to quantify
the mechanism behind these optimal parameters for trapping the energy
in the region between the defects. As a crude approximation, we assume for
the present consideration that energy trapped at the resonator defect sites
is negligible (an assumption partially justified by the limited ability of
each individual defect to trap the energy) and only consider the transmitted
and reflected fractions of the energy from the two defects as follows:
\beq
E_\text{tr}(\epsilon_1,\epsilon_2) = T_1(\epsilon_1)R_1(\epsilon_2)R_2(\epsilon_1)R_3(\epsilon_2).
\label{eq:TrE}
\eeq
The implicit assumption here is that the energy trapped in the region of
interest results from a series of ``favorable'' interactions with the defects,
i.e., an initial transmission from the first defect, followed by a reflection
from the second and then further reflections from both defects
(of which we have only included the first pair in Eq.~(\ref{eq:TrE})).
We first consider
the case of equal-mass defects, $\epsilon_1 = \epsilon_2=\epsilon$, where
$T_1,R_1,R_2$ and $R_3$ are determined in a fitted form from the data obtained in
the top left panel of Figure~\ref{numfig3}. This yields, as a
reasonable approximation,
\beq
\begin{split}
T_1(\epsilon ) &= 0.5630\epsilon^{-0.0110}+0.3345,\\
R_1(\epsilon ) = R_2(\epsilon) = R_3(\epsilon) &= 0.5491\tanh(0.0830 \epsilon)+0.0760,
\end{split}
\label{eq:fits}
\eeq
where both lines were fitted with a relative error $\approx 0.0369$. The plot of
$E_\text{tr}(\epsilon)$ versus the actual, numerically computed trapped energy is shown in
the left panel of Figure~\ref{numfig9} with solid black and red lines, respectively.
The approximation has the right qualitative trend
(especially given our crude assumptions) for the coarser regime
of larger values of $\epsilon$  but breaks down as $\epsilon \rightarrow 0$.
One reason for this discrepancy could be the fact
that we considered the $T$ and $R$ (individual)
fits based on the coarser $\epsilon$ scale. Moreover, the dynamics
of smaller chunks of energy (detached upon collisional events with the
defects from the primary solitary wave) is not adequately captured within
this approximation.

%
We now consider the case where $\epsilon_1 \neq \epsilon_2$ and minimize \eqref{eq:TrE}
using \eqref{eq:fits}. The middle and right panels in Figure~\ref{numfig9} compare the
crude theoretical approximation to the numerical findings. Given the reasonable qualitative
agreement observed, this comparison suggests that the relevant estimation of the trapped
fraction of the energy roughly captures the corresponding main contributions and yields a
semi-analytical handle of some usefulness.

\subsection{Dynamical Trapping of a solitary wave}

In the same spirit as before (of trying to optimize the energy fraction trapped between
the defects), one can envision dynamic protocols that enable capturing a traveling wave
 between two
defects. As just a prototypical example of how to achieve this,
we can consider a dynamic tuning of
the resonator properties that can be performed as the system evolves. Since in realistic experiments
changing the masses is difficult, we need to consider alternative ways to alter the local
properties (and hence the reflection and transmission properties of the defects) on the fly.
Motivated by the recently argued ability to dynamically (in time and/or space) tune elastic
prefactors~\cite{feng}, we consider the possibility of modifying $\kappa$
over time (i.e.,
the prefactor of the linear interaction of the woodpile with its internal resonator), e.g.,
in the form:
\beq
\kappa(t) = k_1 + \left( k_2 - k_1 \right)
\frac{1+ \tanh \left( \frac{t - t_0}{\tau} \right)}{2}.
\label{eq:kappa}
\eeq

\begin{figure}
\centering
\includegraphics[height=.20\textheight, angle=0]{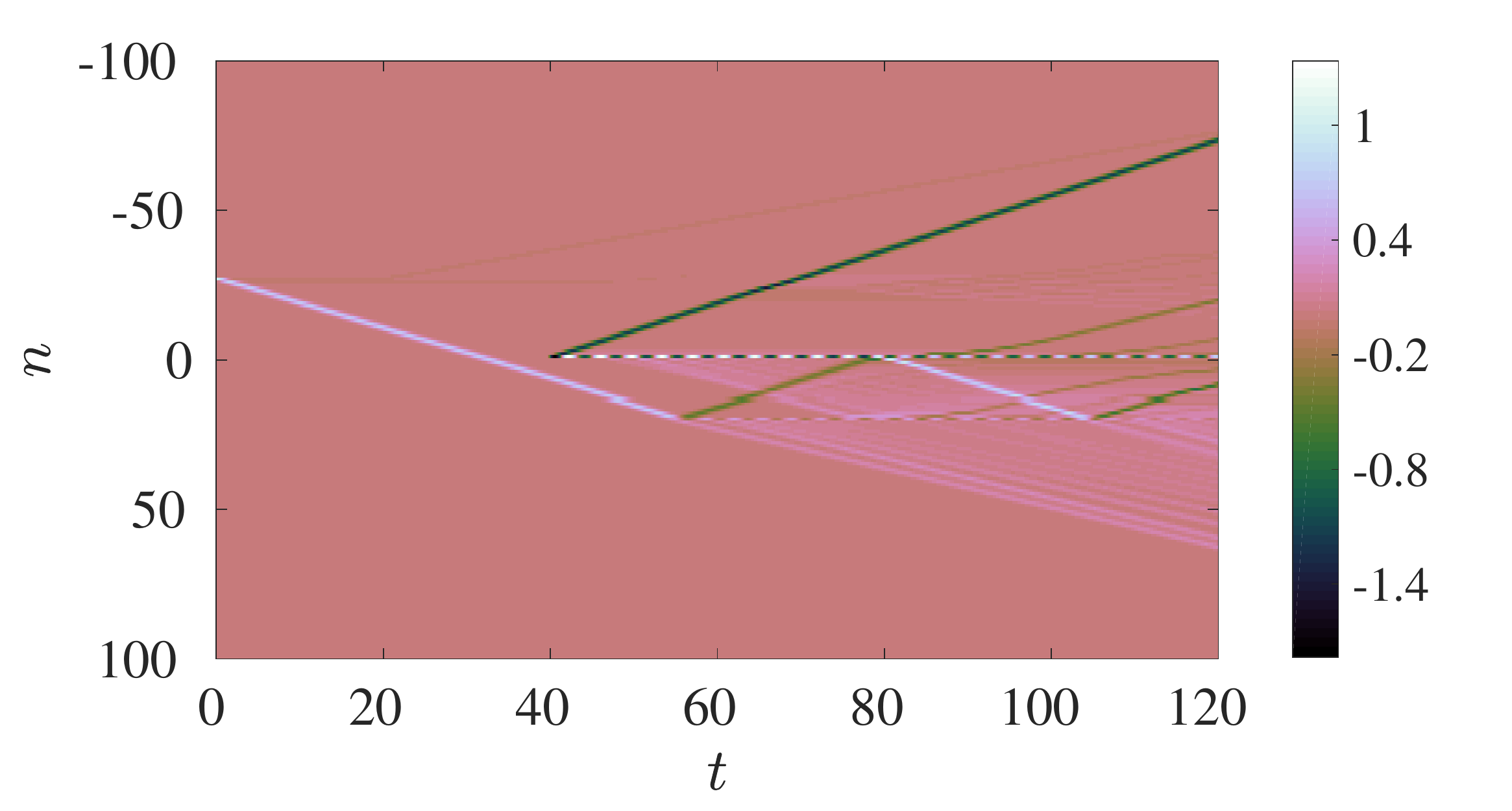}
\caption{(Color online) Space-time velocity plot for the system with two defects of equal mass: $\epsilon_{1}=\epsilon_{2}=10$. 
The first defect, at $d_1=0$, is dynamically swithed on with $\kappa(t)$ given by \eqref{eq:kappa}, where $k_{1}=0$,
$k_{2}=2.5$, $t_{0}=40$, $\tau=0.1$. The second defect,
at $d_2=20$, has fixed $\kappa=2.5$.
Observe that this setting achieves the trapping of a significant fraction of the
solitary wave's energy, in the sense of it bouncing
back and forth between the two defect sites.
\label{numfig11}
}
\end{figure}

This functional form interpolates between $k_1$ and $k_2$, while if considering
dynamics from $t=t_0$ onwards, the interpolation is between $\frac 1 2 (k_1 + k_2)$
at $t=t_{0}$ and $k_2$ is the $\kappa$ as $t\rightarrow\infty$.
In order to thus dynamically trap the solitary wave we envision
the following scenario. We allow the solitary wave to pass from
the first (potential) defect at $d_1=0$ {\it without} having a
defect in that location, i.e., effectively $\kappa(t)=0$ there
when the wave first passes; this way none of the wave's energy is reflected
or transmitted during this first pass. Then, the wave arrives
at $d_2=20$. In the latter location there is a ``fixed'' (not varying
in time) defect with $\kappa=2.5$. Notice that for both locations,
we have selected a mass of $\epsilon_1=\epsilon_2=10$. Once the wave arrives
at $d_2=20$, we can observe in its dynamical evolution shown in
Figure~\ref{numfig11} that it gets chiefly
reflected.

During the time frame when the wave moves from $d_1=0$ to $d_2=20$,
the dynamical defect at $d_1=0$ is put in place. In particular,
we use Eq.~(\ref{eq:kappa}) with $k_{1}=0$, $k_{2}=2.5$, $t_{0}=40$, $\tau=0.1$,
i.e., a defect with $\kappa=2.5$ arises at this location within a short time frame around
$t \approx 40$. Unfortunately, due to
the traveling front nature of the wave, this causes a trapping and reflection
at the location of the wave (observed
in Figure~\ref{numfig11}),
however, this is mostly inconsequential in connection to the
propagation of the wave. The most adverse side effect of this is that
a small fraction of energy created by the ``raising'' of the defect
at $d_1=0$ propagates inside the region between $d_1$ and $d_2$
and affects both (weakly) the motion of the wave and (also weakly
but nontrivially) the amount of trapped energy in this region.
Importantly, once this nontrivial defect at $d_1=0$ has been dynamically
raised, it causes the wave to subsequently be chiefly reflected both
at $d_1=0$ and at the fixed defect at $d_2=20$ with its energy remaining
mainly  trapped in the region between the two defects. This dynamical
emergence of a defect clearly achieves the confinement of the wave's largest
energy fraction within the desired region.
One can naturally envision multiple alternative scenarios leading to such
a confinement, yet we believe that this simple proof of principle illustrates
the main idea and can motivate further studies along this vein.

\section{Conclusions and Future Challenges}
In the present work, we have considered a nonlinear granular chain with multiple MwM
defects that can be experimentally tested in some parameter regimes. Using a woodpile
elastic lattice as our experimental motivation, we illustrated how one can realize and
probe such a system for the cases of one and two defects represented by longer rods in the
orthogonally stacked chain. Laser Doppler vibrometry then enabled us to measure the (kinetic)
energy transmitted, reflected and trapped at this ``defective'' region. We then turned to
a series of theoretical and numerical considerations. We examined how the fraction of trapped
(and transmitted/reflected) energy scales as a function of the size of the region bearing the
resonators. We also explored the possibility of stiffer nonlinear internal resonators and
determined that the near-linear ones enable the highest energy trapping. We considered the
variation of the masses of the resonators and were able to numerically optimize the trapped
region, as well as obtain a qualitative understanding of this optimization on the basis of
transmissions and reflections of the principal traveling wave within the system. Finally, we
proposed
dynamical scenarios of variable elastic properties and utilized them to further
enhance the potential of trapping energy within the region enclosed by our ``defects''.

This study opens numerous directions for the potential of directing and manipulating energy
in such elastic woodpile lattices. The tunability of the rod lengths and possibly also of
the elastic constants provides a large potential for considering defect regions of different
sizes and properties in a highly tractable and controllable experimental setting. 
Extending such considerations to two-dimensional woodpile lattices and achieving the
steering, and channeling of the energy in a controllable
fashion, possibly reminiscent of
analogous propositions in optics~\cite{eugenieva}, may be of particular interest for future
theoretical and experimental work.

\begin{acknowledgments}

J.Y. and E.K. are grateful for the support from the National Science Foundation through the grant 
CAREER-1553202. E.K. acknowledges the support from the National Research Foundation of Korea (NRF) 
grant funded by the Korea government (MSPI, No. 2017R1C1B5018136). J.Y. and P.G.K. also acknowledge 
the support from the ARO (W911NF-15-1-0604). P.G.K. acknowledges that this paper was made possible 
by NPRP grant $\#~[8-764-1-160]$ from the Qatar National Research Fund (a member of Qatar Foundation). 
The findings achieved herein are solely the responsibility of the authors. P.G.K. also acknowledges 
support from the National Science Foundation under Grants DMS-1312856 and PHY-1602994, the Alexander 
von Humboldt Foundation, the Stavros Niarchos Foundation via the Greek Diaspora Fellowship Program.
He also acknowledges useful discussions with Alex Vakakis. X.H and S.H gratefully acknowledge funding 
from the Department of Mathematics and Statistics (UMass) under the ``Research Experience for Undergraduates'' 
(REU) program. The work of A.V. was supported by the U.S. National Science Foundation through the grant 
DMS-1506904.

\end{acknowledgments}

\end{document}